\documentclass[%
reprint,
superscriptaddress,
amsmath,amssymb,
aps,
prb,
floatfix,
longbibliography
]{revtex4-2}

\usepackage{physics}
\usepackage{graphicx}
\usepackage{dcolumn}
\usepackage{bm}
\usepackage{graphicx} 
\usepackage[dvipsnames]{xcolor}
\usepackage[bookmarks=false,linkcolor=Orange,urlcolor=MidnightBlue,colorlinks,citecolor=Maroon]{hyperref}
\usepackage[caption=false]{subfig}

\NewDocumentEnvironment{FullWidth}{ +b }{
    \twocolumn[{#1}]%
}{}
\begin{document}

\title{Nonlinear optical responses and quantum geometry in rhombohedral trilayer graphene}
\author{Abigail Postlewaite}
\affiliation{Department of Physics, Northeastern University, Boston, Massachusetts 02115, USA}

\author{Arpit Raj}
\affiliation{Department of Physics, Northeastern University, Boston, Massachusetts 02115, USA}

\author{Swati Chaudhary}
\affiliation{Department of Physics, The University of Texas at Austin, Austin, Texas 78712, USA}
\affiliation{Department of Physics, Northeastern University, Boston, Massachusetts 02115, USA}
\affiliation{Department of Physics, Massachusetts Institute of Technology, Cambridge, Massachusetts 02139, USA}

\author{Gregory A. Fiete}
\affiliation{Department of Physics, Northeastern University, Boston, Massachusetts 02115, USA}
\affiliation{Department of Physics, Massachusetts Institute of Technology, Cambridge, Massachusetts 02139, USA}

\begin{abstract}
We theoretically study the nonlinear optical response of ABC trilayer graphene with inversion symmetry broken by the application of a displacement field perpendicular to the trilayer. We show that rhombohedral trilayer graphene exhibits a large bulk photovoltaic effect arising from a DC shift current response. The conductivity of the trilayer contains features similar to AB bilayer graphene as well as features distinct from AB bilayer graphene. The new features of ABC trilayer graphene relative to AB bilayer graphene arise from the quantum geometric features of the electronic band structure and can be tuned by varying the displacement field. We focus on a regime of displacement field where certain band gaps close and reopen away from the charge neutrality point leading to drastic changes in the quantum geometric structure in momentum space, a feature characteristic of the trilayer graphene band structure. These features manifest as a sign change in shift current conductivity in a certain frequency window and can thus serve as a probe of quantum geometry. 
\end{abstract}

\maketitle

\section{Introduction}

Graphene has garnered much attention in condensed matter physics due to the properties of its band structure with Dirac points at the corners of the first Brillouin zone that are crucial for its transport and optical properties, as well as electron interaction instabilities such as superconductivity and magnetism. The characteristics of graphene's band structure result in enhanced conductivity as well as topological phenomena such as the integer quantum Hall effect at half-integer fillings and quantized anomalous Hall effect, with numerous applications in materials science and spintronics \cite{ferrari2015science,PhysRevB.90.094516,Novoselov2005-fx,PhysRevLett.114.016603,PhysRevLett.128.026403,SOLTANIZANJANI2021104326,romagnoli2018graphene,castro2009electronic,das2011electronic}. 

Layered configurations of graphene have also emerged as a topic of interest within the field~\cite{mccann2013electronic,guerrero2022rhombohedral,han2023orbital,lui2011observation,yacoby2011tri,zhou2021half,shi2020electronic,Ghazaryan2023,Huang2023,Koh2024,zhou2021superconductivity,chen2019evidence}, with layered graphene retaining its two-dimensional characteristics up to roughly 10-layer systems \cite{GeimA.K2009GSaP, PhysRevB.107.104502,Murata_2020}. Layered graphene is roughly categorized into monolayer, bilayer and few-layered (n = 3-10) systems. The bandstructures of layered graphene systems rapidly evolve as the number of layers increase~\cite{Nilsson2008}, and as a function of stacking order \cite{mccann2013electronic}. While monolayer graphene is characterized by the familiar Dirac cones, the bandstructure of bilayer graphene hosts a parabolic dispersion for AB stacking \cite{mccann2013electronic,rozhkov2016electronic}. Trilayer ABC graphene, further diverging from the bandstructure of the original monolayer configuration, displays a cubic dispersion~\cite{yacoby2011tri,PhysRevB.82.035409}. The increased degrees of freedom offered by additional layers allow novel phenomena to emerge without sacrificing the simplicity of the two-dimensional character of the material, making few-layer graphene systems a source of rich theoretical and experimental promise \cite{PhysRevB.101.165437,PhysRevLett.124.166601,Park2022,Park2021,yacoby2011tri, PhysRevB.80.235402}. 

The band structure of layered graphene systems can be tuned drastically via twisting of the layers relative to one another, manipulation of the stacking order, and by gating~\cite{aoki2007dependence,bao2017stacking,zhang2009direct,lui2011observation,Liang2020,zhou2022isospin}. Untwisted bilayer graphene systems display either AA  or AB stacking order~\cite{birowska2011van,Liang2020}, while stable allotropes of trilayer graphene are limited to either ABC or ABA configurations \cite{PhysRevB.84.161408}. The transport properties of both bilayer and trilayer graphene systems are markedly different depending on stacking configuration \cite{Bao2011,PhysRevB.108.235401}. In ABC trilayer graphene, application of a perpendicular electric field--referred to as a displacement field--gaps out the band structure, while the same treatment of ABA trilayer graphene causes the band overlap to increase. Thus, ABA trilayer graphene remains metallic even with application of a displacement field \cite{PhysRevB.84.161408, PhysRevB.84.125455,doi:10.1021/acs.chemmater.0c01145}. ABC trilayer graphene also admits a more robust ferromagnetic state than AB bilayer graphene or ABA trilayer graphene \cite{PhysRevB.87.115414}.

In addition to highly tunable band structures, multilayered graphene systems have proven to be an ideal platform to study electron-electron interactions and quantum geometric effects. Quantum geometric effects in the electronic eigenstates lead to strong signatures in transport and optical responses, and in some cases can even promote unconventional superconducting states \cite{Törmä2022,doi:10.1021/acs.nanolett.0c04494}. Photocurrent studies in multilayer graphene systems have also gained significant attention~\cite{ma2023photocurrent,ma2021topology,kumar2024terahertzphotocurrentprobequantum,Sunku2021} as they provide direct access to symmetry breaking and quantum geometric features. In particular, many  graphene multilayer systems have been predicted to exhibit a significant bulk photovoltaic effect (BPVE)~\cite{kumar2024terahertzphotocurrentprobequantum,chaudhary2022shift,Kaplan2022TBG,zheng2023gate,chen2024enhancing,Koppens2014} arising from a second-order optical process. 

The leading second-order responses for the BPVE include the shift and injection currents, both of which share quantum geometric origins and  require broken inversion symmetry. Shift current arises due to the shift in the center of electron wavepackets upon making a band transition~\cite{OrensteinJ2021TaSo,Parker19}, which is related to a change in the Berry connection and can be produced with linearly polarized light. On the other hand, injection current requires circularly polarized light in time-reversal preserving systems \cite{de2017quantized,Parker19}. Both of these effects originate from different quantum geometric quantities.

Quantum geometry can be described by the quantum geometric tensor, from which one can obtain the Berry curvature and quantum metric, among other geometric objects~\cite{resta2011insulating}. These entities have been found to be useful in describing static electronic responses involving a single quantum state, or in certain limits in which a system can be adequately described by a two-band model \cite{ahn2022riemannian}. However, resonant optical transitions arising from interband effects are less easily quantified by such metrics. Thus, it is an active area of interest to determine to what extent the features in nonlinear optical responses can be correlated with quantum geometric properties of a system~\cite{Neufeld2023,Ma2022,Bao2022,PhysRevB.109.184514,ghosh2024probing,PhysRevX.14.011052}. It was determined that second-harmonic generation, a resonant optical response characterized by frequency doubling of the incident electromagnetic radiation, could be used as a probe of Berry curvature, quantum metric and quantum geometric connection \cite{bhalla2022resonant}. Third-order responses were also studied and correlated with the Riemannian curvature tensor \cite{ahn2022riemannian}, and the superconducting quasi-particle Berry curvature was found to play a key role in the nonlinear optical response of a noncentrosymmetric superconductor \cite{PhysRevB.107.024513}. 

Further work remains to be done in order to determine the utility of nonlinear optical responses as a probe of quantum geometry in few-layer graphene systems. We address this gap by studying the shift current response of an ABC trilayer graphene system with inversion symmetry broken via application of a tunable perpendicular electric (displacement) field for signatures of quantum geometry. We do so by calculating the DC shift current response in a low-frequency regime and examining the results for correlations between the shift conductivity and the quantum metric and Berry curvature, derived from the real and imaginary parts of the quantum geometric tensor, respectively.  We find signatures of the quantum geometric tensor reflected in the nonlinear optical response of ABC trilayer graphene.

Our paper is organized as follows. Section~\ref{sec:model} describes the model Hamiltonians for the bilayer and trilayer graphene systems, including the displacement field, intra- and interlayer hoppings. In Section~\ref{sec:nlor}, we provide the necessary formalism for calculating the shift conductivity and discuss the symmetry constraints on a nonzero response along with the symmetry relations between various components of the second order conductivity tensor. We present and discuss our numerical calculations of the optical conductivity (shift current) in Section~\ref{sec:results} and briefly discuss the quantum geometric quantities considered, with further details and numerical results given in the appendices. We finish with our primary conclusions and an outlook for future work in Section~\ref{sec:conc}.

\section{Model}
\label{sec:model}
We study a trilayer graphene system with ABC stacking order, characterized by a tight-binding model following the work of Ref.\cite{PhysRevB.88.075408} with the Hamiltonian
\begin{widetext}
\begin{equation} \label{Habc}
    H^\mathrm{ABC}(\vb{k}) = -
    \begin{bmatrix}
        -\Delta-\Delta^\prime/2 & \gamma_0f(\vb{k}) & 0 & \gamma_3 f^*(\vb{k})  & 0 & \gamma_2 \\
        \gamma_0 f^*(\vb{k}) & -\Delta+\Delta^\prime/2 & \gamma_1 & 0 & 0 & 0 \\
        0 & \gamma_1 & \Delta^\prime/2 & \gamma_0f(\vb{k}) & 0 & \gamma_3 f^*(\vb{k}) \\
        \gamma_3 f(\vb{k})& 0 & \gamma_0 f^*(\vb{k}) & \Delta^\prime/2 & \gamma_1 & 0 \\
        0 & 0 & 0 & \gamma_1 & \Delta+\Delta^\prime/2 & \gamma_0 f(\vb{k}) \\
        \gamma_2 & 0 & \gamma_3 f(\vb{k}) & 0 & \gamma_0f^*(\vb{k}) & \Delta-\Delta^\prime/2
    \end{bmatrix},
\end{equation}
where 
\begin{equation}
    f(\vb{k}) = e^{ik_ya/\sqrt{3}}\bigg[1+2e^{-3ik_ya/2\sqrt{3}}\cos\bigg(\frac{k_xa}{2}\bigg)\bigg],
\end{equation}
\end{widetext}

using the basis $\left(\psi_{A_1,\vb{k}},\psi_{B_1,\vb{k}},\psi_{A2,\vb{k}},\psi_{B_2,\vb{k}},\psi_{A_3,\vb{k}},\psi_{B_3,\vb{k}}\right)$
where $a$ is the lattice constant of the honeycomb lattice and $^*$ indicates the complex conjugate. 

\begin{figure}
    \centering
    \includegraphics[scale=0.5]{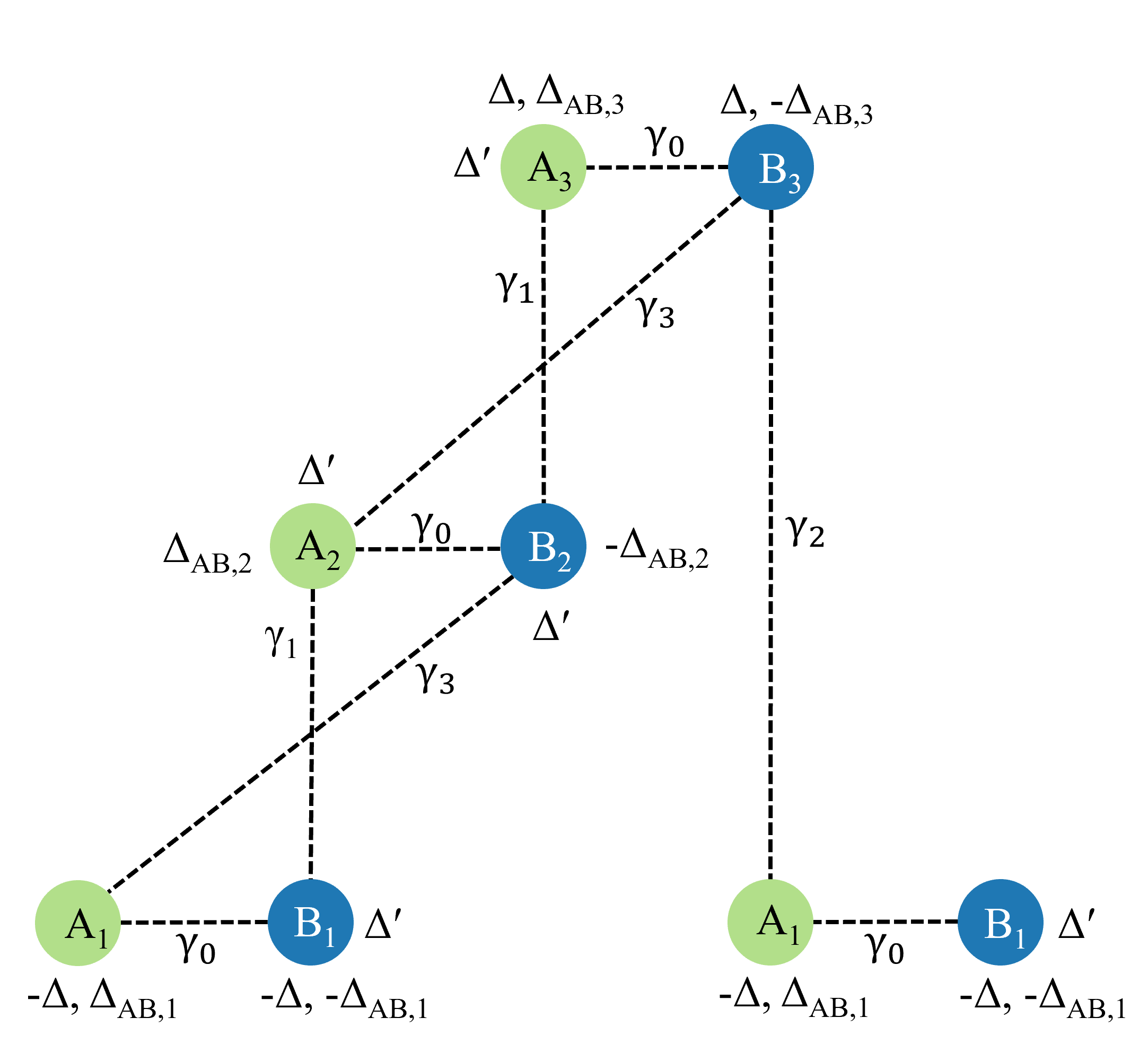}
    \caption{Schematic diagram of ABC trilayer graphene. Shown is a cross-section of a stacking of graphene layers with A and B sublattice sites labeled.  The parameter $\gamma_0$ describes intralayer first neighbor hopping, $\gamma_1$ interlayer A-B site hopping when the two atoms are "on top" of each other, $\gamma_2$ the A-B site hopping between the top and bottom layers of the trilayers, and $\gamma_3$ the interlayer hopping between A and B sites not on top of one another. $\Delta$ is the displacement field, $\Delta'$ a dimerization potential taken to be zero in our study, and $\Delta_{AB}$ a sub-lattice offset term, which is also taken to be zero in our study.}
    \label{ABC_hoppings}
\end{figure}

\begin{figure}[h!t!]
    \centering
    \includegraphics[scale=0.5]{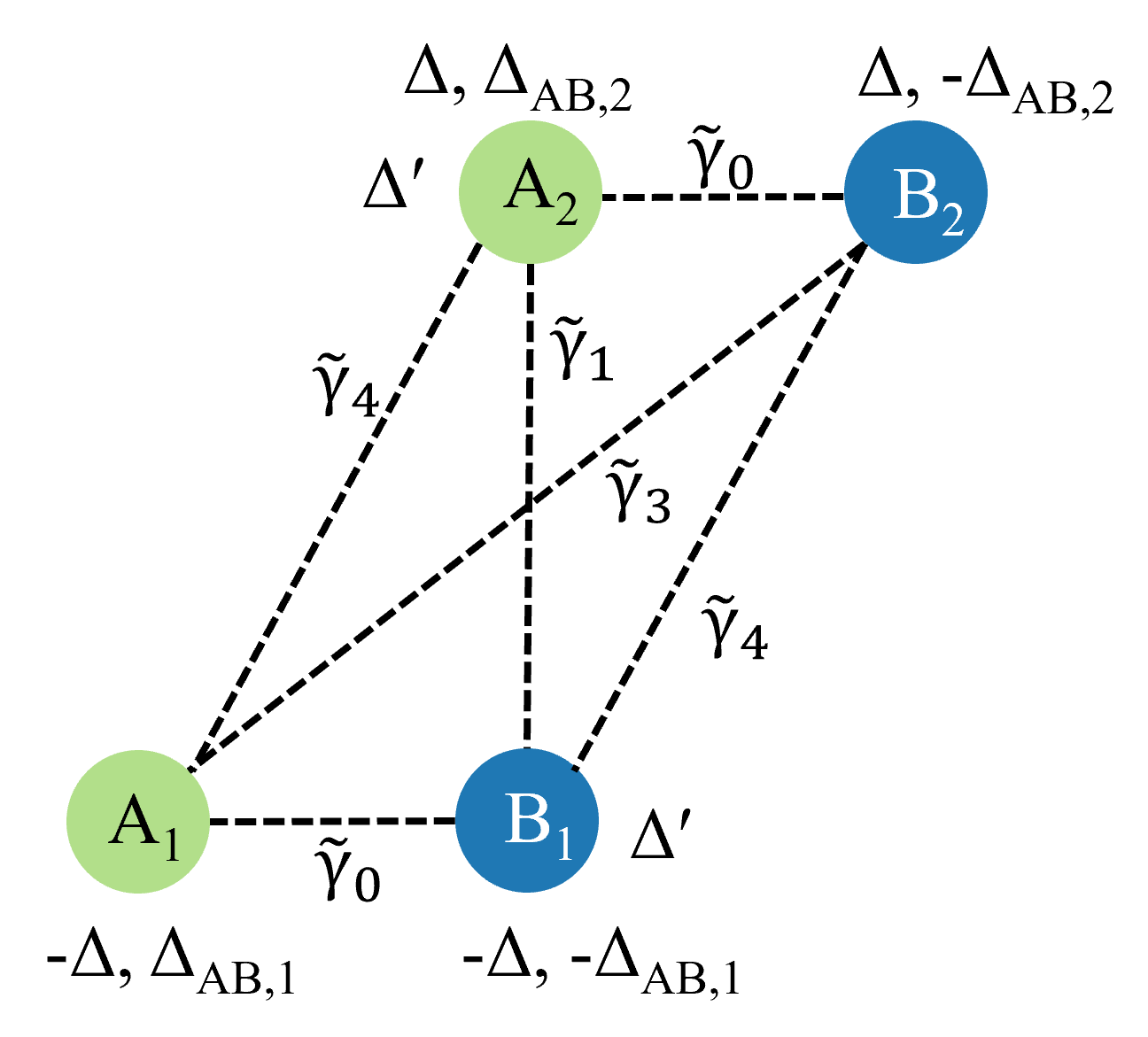}
    \caption{Schematic diagram of AB bilayer graphene. Shown is a cross-section of a stacking of graphene layers with A and B sublattice sites labelled.  The parameter $\tilde \gamma_0$ discribes intralayer first neighbor hopping, $\tilde \gamma_1$ interlayer A-B site hopping when the two atoms are ``on top" of each other, $\tilde \gamma_3$ the A-B site hopping between the top and bottom layers of the bilayer, and $\tilde \gamma_4$ the interlayer hopping between A and A (or B and B) sites. Here $\Delta$ is the displacement field, $\Delta'$ a dimerization potential taken to be zero in our study, and $\Delta_{AB}$ a sub-lattice offset term, which is also taken to be zero in our study.}
    \label{AB_hoppings}
\end{figure}

In $H^{\text{ABC}}(\vb{k})$ in Eq.\eqref{Habc}, $\gamma_i$ are hopping parameters characterizing the coupling between lattice sites with $a = 2.46\mathrm{\r{A}}$, $\gamma_0 = 3.12\,\text{eV}$, $\gamma_1 = 0.377\,\text{eV}$, $\gamma_2 = 0.01\,\text{eV}$, and $\gamma_3 = 0.3\,\text{eV}$, shown in Fig.~\ref{ABC_hoppings}.  The displacement field characterized by $\Delta$ breaks inversion symmetry in the system, while $\Delta^\prime$ gives the dimerization potential, which is taken to be zero in this analysis. The schematic given in Fig.~\ref{ABC_hoppings} includes the additional parameter $\Delta_{AB}$, a sub-lattice offset term, which is also taken to be zero in our study. 
A brief analysis of AB bilayer graphene is used for comparison with ABC trilayer graphene, using a tight-binding Hamiltonian given by 
\begin{widetext}
\begin{equation} \label{Hab}
H^{AB}(\vb{k}) = 
    \begin{bmatrix}
            -\Delta - \Delta^\prime/2 & \tilde\gamma_0 f(\vb{k})& \tilde\gamma_4 f(\vb{k}) & \tilde\gamma_3 f^*(\vb{k}) \\
            \tilde\gamma_0 f^*(\vb{k}) & -\Delta + \Delta^\prime/2 & \tilde\gamma_1 & \tilde\gamma_4 f(\vb{k}) \\
            \tilde\gamma_4 f^*(\vb{k}) & \tilde\gamma_1 & \Delta + \Delta^\prime/2 & \tilde\gamma_0 f(\vb{k}) \\
            \tilde\gamma_3 f(\vb{k}) & \tilde\gamma_4 f^*(\vb{k}) & \tilde\gamma_0 f^*(\vb{k}) & \Delta - \Delta^\prime/2 
        \end{bmatrix},
\end{equation}
\end{widetext}

as found in Ref.\cite{zheng2023gate}, with $\tilde\gamma_0 = -3.16\,\text{eV}$, $\tilde\gamma_1 = 0.381\,\text{eV}$, $\tilde\gamma_3 = -0.38\,\text{eV}$, and $\tilde\gamma_4 = 0.14\,\text{eV}$, and $\Delta^\prime = 0\,\text{eV}$. The lattice configuration of AB bilayer graphene can be seen in Fig.~\ref{AB_hoppings}, which also includes $\Delta^\prime$ and $\Delta_{AB,n}$ parameters taken to be zero in this analysis.
\begin{figure}[ht!]
    \centering
    \includegraphics[scale=0.27]{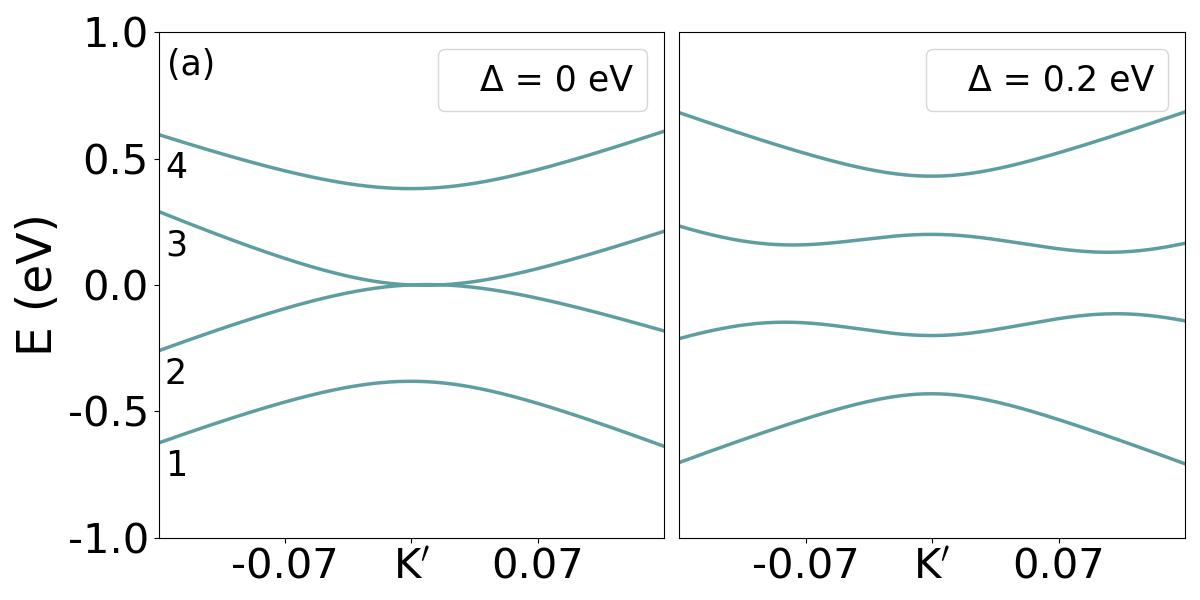}
    \includegraphics[scale=0.27]{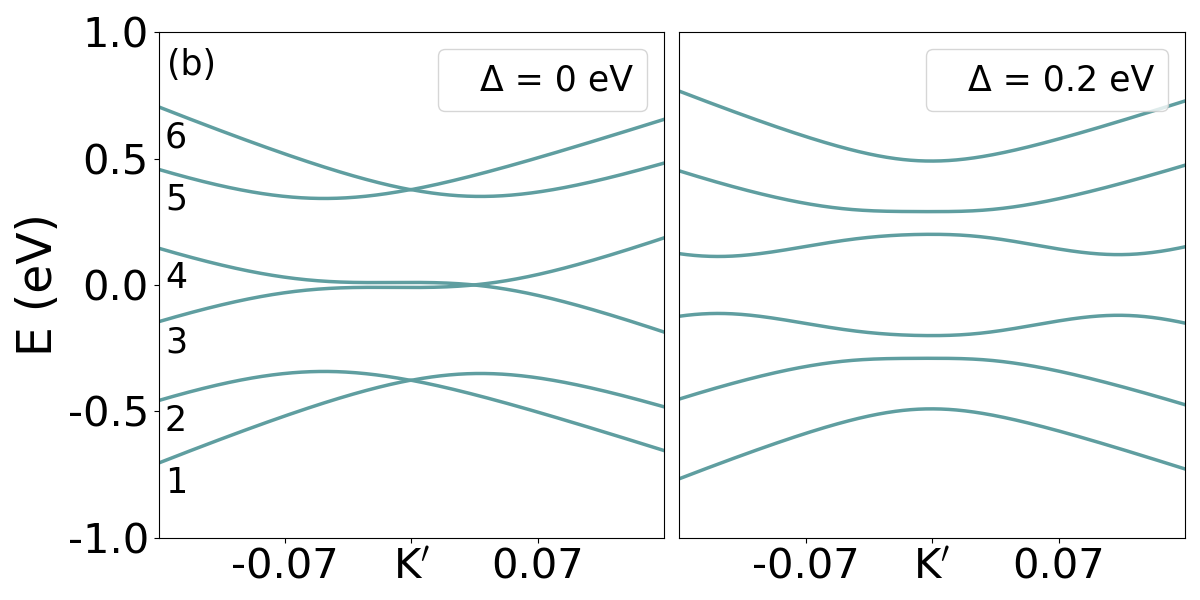}
    \caption{(a) Band structure of AB bilayer graphene for $\Delta = 0,\,0.2\,\text{eV}$. Energy bands are labeled with indices $1-4$. (b) Band structure of ABC trilayer graphene for $\Delta=0,\,0.2\,\text{eV}$. Energy bands are labeled with indices $1-6$. Note that a gap opens between all bands for $\Delta\neq0$.}
    \label{figbandstructure}
\end{figure}

In both AB bilayer graphene and ABC trilayer graphene, $\Delta = 0\,\text{eV}$ corresponds to a gapless system with inversion and $C_3$ symmetries. Inversion symmetry prohibits a nonzero nonlinear optical response, and therefore must be broken for a nonlinear response to appear. The parameter $\Delta$ characterizing the displacement field (electric field perpendicular to the planes of the lattices) opens a tunable gap in the system, as can be seen in Fig.~\ref{figbandstructure}. 

\section{Nonlinear Optical Response}
\label{sec:nlor}
We calculate the shift current, a second-order DC response, focusing on the low-frequency response ($\omega < 1\,\text{eV}$) as the displacement field varies, parameterized by $\Delta$. We evaluate the shift current conductivity using the expression \cite{Cook2017}, 
\begin{equation}
    \sigma^{a;bc}(\omega) = \frac{2g_s\pi e^3}{\hbar^2} \int \frac{dk^2}{(2\pi)^2}\sum\limits_{n,m}f_{nm}I_{nm}^{a;bc}\delta(\omega_{nm}-\omega)
    \label{eq:sigma}
\end{equation}
with $e$ the electron charge and $g_s=2$ accounts for the spin degeneracy. In Eq.\eqref{eq:sigma}, $f_{nm} = f_n-f_m$ for the Fermi occupations $f_n = 1/(1+e^{E_n/{k_BT}})$ where $T$ is the temperature, $k_B$ is Boltzmann's constant, $E_n$ is the energy of the $n^{th}$ state, $\delta(\omega_{nm}-\omega)$ is a Dirac delta function approximated by a Lorentzian with broadening parameter $\Gamma = 0.01\,\text{eV}$, $\omega_{nm} = (E_n-E_m)/\hbar$ (where we take $\hbar = 1$), and the integrand $I_{nm}^{a;bc}$ for conductivity tensor elements $\sigma^{a;bb}$ is given by 
\begin{equation}
    I_{nm}^{a;bb} = \mathrm{Im}(r^b_{mn}r^b_{nm;a}),
\end{equation}
where the matrix elements $r^a_{nm}$, $r^a_{nm;b}$ are the interband matrix elements of the position operator $r^a_{nm}=i\bra{n} \partial_{k_a} \ket{m}$, and its generalized derivative given by 
\begin{equation}
    r^b_{nm;a} = \left[\partial_{k_a}-i(\xi^a_{nn}-\xi^a_{mm})\right]r^b_{nm},
\end{equation}
respectively. 

Both $r^a_{nm}$ and $r^a_{nm;b}$ can be computed from velocity matrix elements $v^a_{nm}=\langle n | \partial_{k_a}H|m\rangle$ and eigenenergies $E_n$ of the Hamiltonian $H(\mathbf{k})$ via the following expressions: 
\begin{equation}
    r^a_{nm} = \frac{v^a_{nm}}{i\omega_{nm}}, \quad n\neq m
\end{equation}
\begin{equation}
\begin{split}
    r^a_{nm;b} &=  \frac{i}{\omega_{nm}}\bigg[\frac{v^a_{nm}\Delta^b_{nm}+v^b_{nm}\Delta^a_{nm}}{\omega_{nm}} - w^{ab}_{nm} \\
    &\quad + \sum\limits_{p\neq n,m} \left(\frac{v^a_{np}v^b_{pm}}{\omega_{pm}}-\frac{v^b_{np}v^a_{pm}}{\omega_{np}}\right) \bigg], \quad n\neq m
\end{split}
\end{equation}
where $w^{ab}_{nm} = \langle n | \partial_{k_a}\partial_{k_b} H | m \rangle$, and $\Delta^a_{nm} = v^a_{nn}-v^a_{mm}$. The three-fold rotation symmetry in plane of the material, $C_3$, forbids nonzero $-\sigma^{x;xx} = \sigma^{x;yy} = \sigma^{y:xy} = \sigma^{y;yx}$ and requires that $-\sigma^{y;yy} = \sigma^{y;xx}= \sigma^{x;yx}= \sigma^{x;xy}$. Due to the restrictions imposed by the above symmetries, we limit our studies to the component $\sigma^{y;yy}(\omega)$ for signatures corresponding to quantum geometric features of the system.

The interband matrix elements $r^b_{nm}$ can also be used in constructing the quantum geometric tensor $Q^n_{ab}$,

from which the quantum metric $g_{ba} \equiv \Re [Q_{ab}^n]$ and the Berry curvature $F_{ba} \equiv -2\Im [Q_{ab}^n]$ can be calculated (see Appendix \ref{Qabnapp}). Another geometric quantity, the Hermitian connection, $C_{cba}^{mn}$,  is constructed from $r^b_{nm}$ and its generalized derivative $r^b_{nm;a}$. Its real and imaginary parts can be used to calculate quantities known as the metric connection and symplectic connection (see Appendix \ref{Cyyyapp}.)

\section{Results}
\label{sec:results}
The shift-current conductivity is determined by many factors: the quantum geometric quantities, the joint density of states (JDOS), the electronic band structure, and the band occupancies. One of the most remarkable features of multilayered graphene systems is that their electronic properties can be tuned easily by applying a displacement field or gate voltage. Here, we study the variation of the shift-current conductivity with displacement field at the charge-neutrality point. 

A displacement field along the axis perpendicular to the lattice layers breaks inversion symmetry in both the AB and ABC graphene systems. It can be seen in the band structure that the presence of a nonzero displacement field opens a gap in the AB and ABC trilayer graphene systems, as observed in Fig.~\ref{figbandstructure}. The displacement field denoted by parameter $\Delta$ can be tuned via a perpendicular electric field. In our calculations, we vary the parameter $\Delta$, which corresponds to tuning of the displacement field, and study changes in the conductivity $\sigma^{y;yy}(\omega)$, focusing on the  low-frequency regime ($\omega<1\,\text{eV}$). 

\begin{figure}[th!]
    \centering
    \includegraphics[scale=0.27]{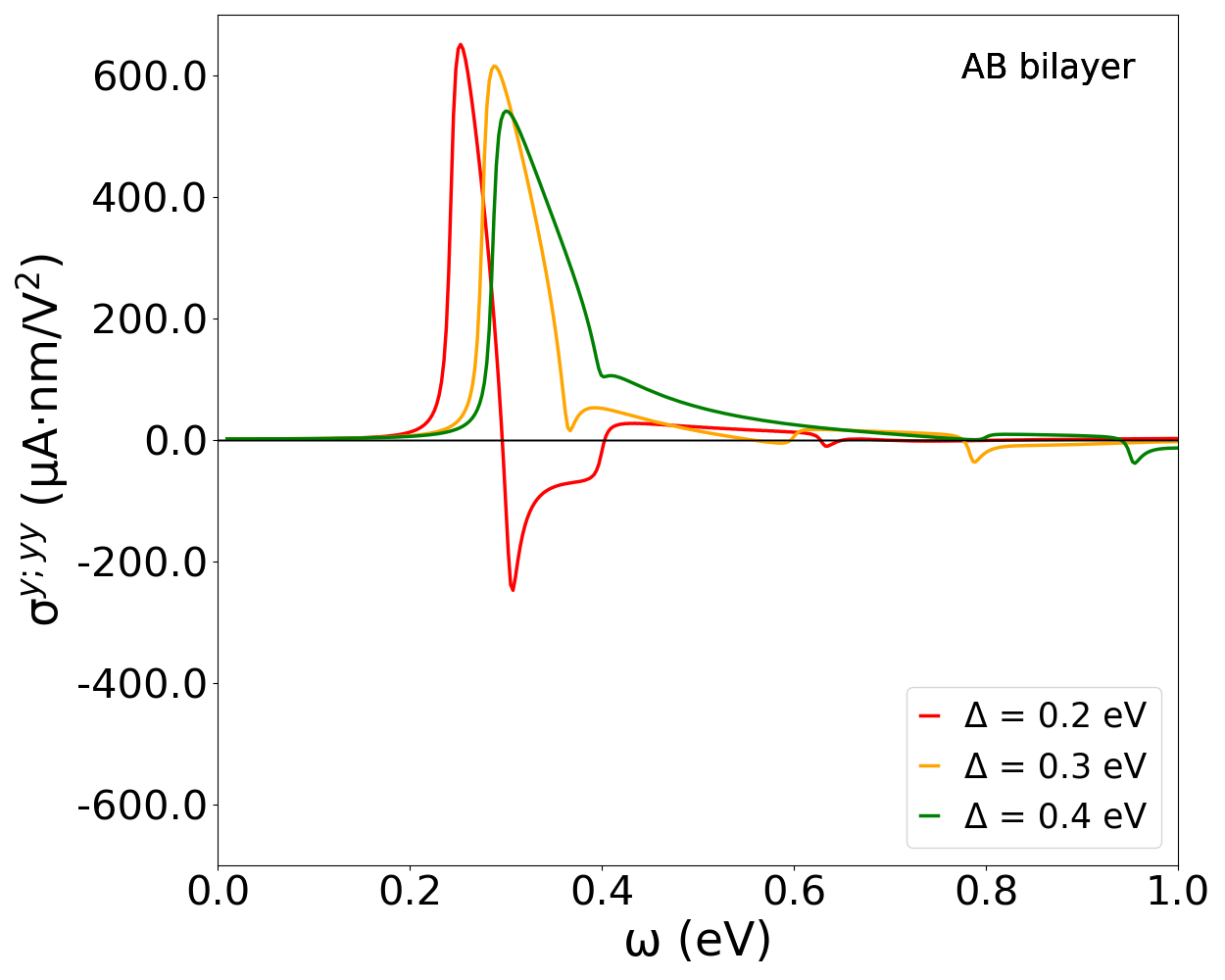}
    \vspace{0cm}
    \includegraphics[scale=0.27]{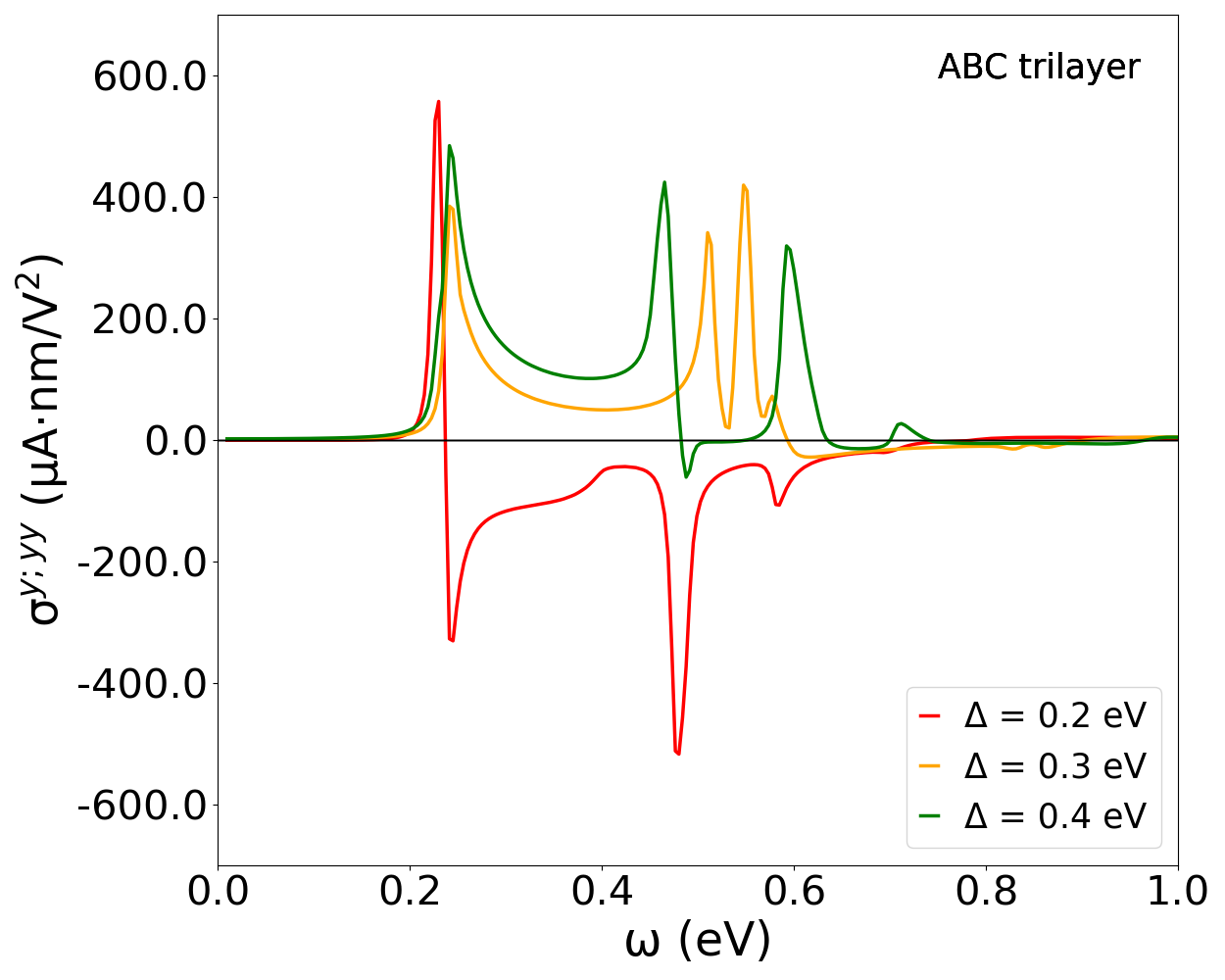}
    \caption{The nonlinear conductivity tensor component $\sigma^{y;yy}$ for AB bilayer graphene (top) and ABC trilayer graphene (bottom) at $\Delta = 0.2,\, 0.3,\, 0.4\,\text{eV}$.  In Fig.\ref{conductivity-colors} we focus the evolution of the conductivity with displacement field in a narrower range of frequencies and displacement field where a band transition occurs. }
    \label{figsigmaAB_ABC}
\end{figure}
We first compare the conductivities of AB bilayer graphene and ABC trilayer graphene within this low-frequency regime for a coarse range of values of $\Delta$. The conductivity was computed for both systems for $\Delta=0.2\,\text{eV}$, $0.3\,\text{eV}$, and $0.4\,\text{eV}$. See Fig.\ref{figsigmaAB_ABC}.

\begin{figure*}[ht!]
    \centering
    \includegraphics[scale=0.27]{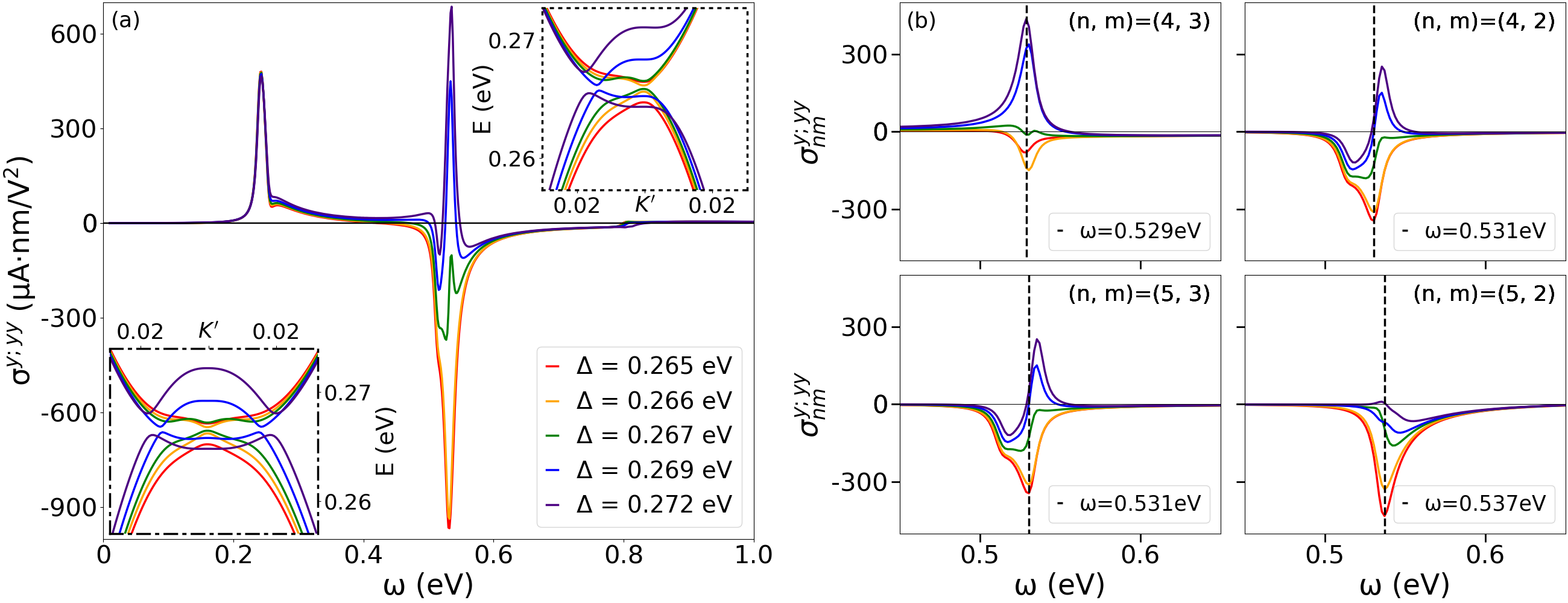}
    \caption{ (a) Full conductivity $\sigma^{y;yy}(\omega)$ , (b) band-resolved conductivity $\sigma^{y;yy}_{nm}(\omega)$ and variation of bands 4, 5 (insets) for a range of displacement field $\Delta$ values from $0.265-0.272\,\text{eV}$. Dashed inset: $\Gamma-\text{K}^\prime-\text{M}$ path through the FBZ. Dash-dot inset: $\Gamma-\text{K}^\prime-\text{K}$ path through the FBZ.}
    \label{conductivity-colors}
\end{figure*}

\begin{figure}[h!]
    \centering
    \includegraphics[scale=0.27]{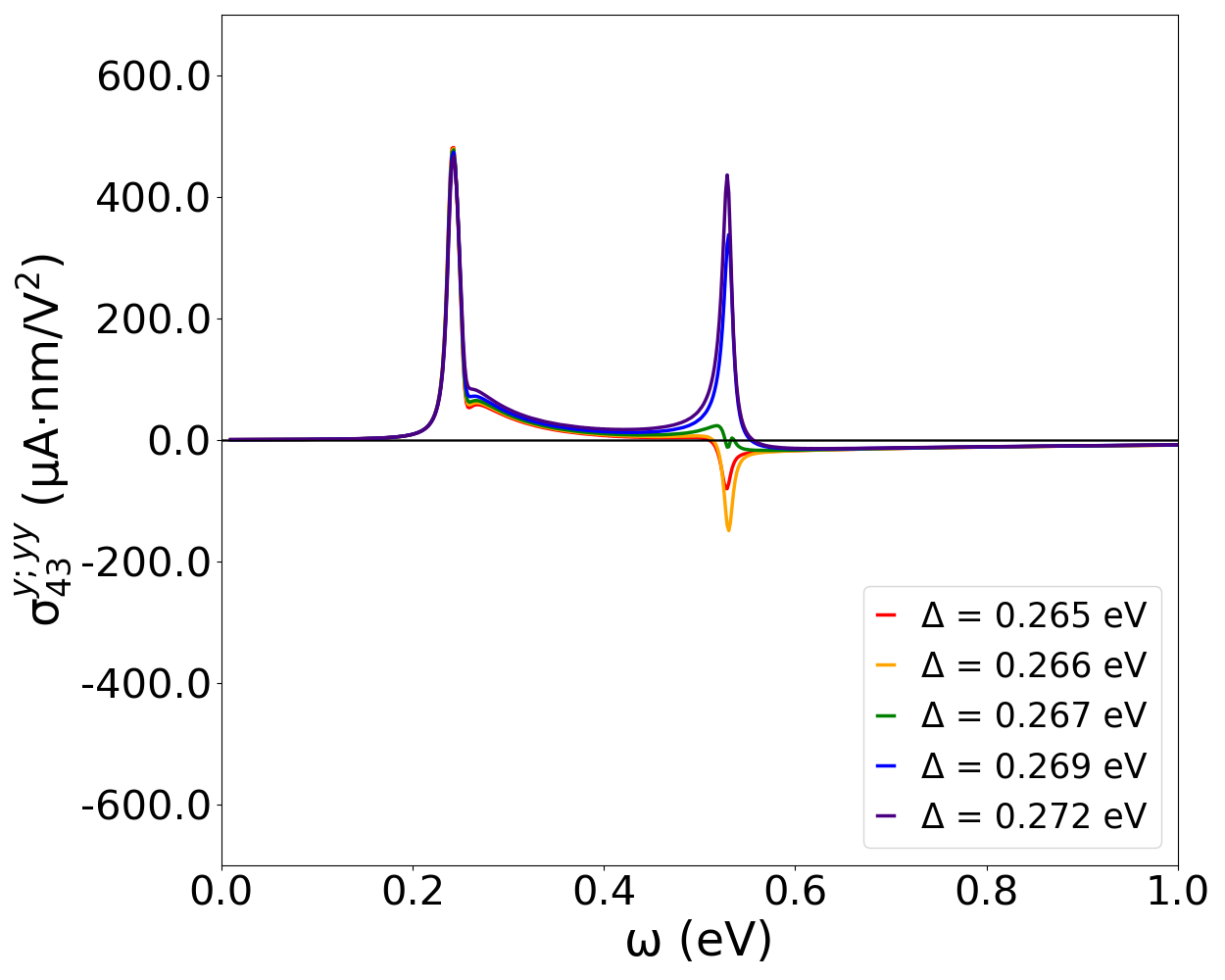}
    \includegraphics[scale=0.27]{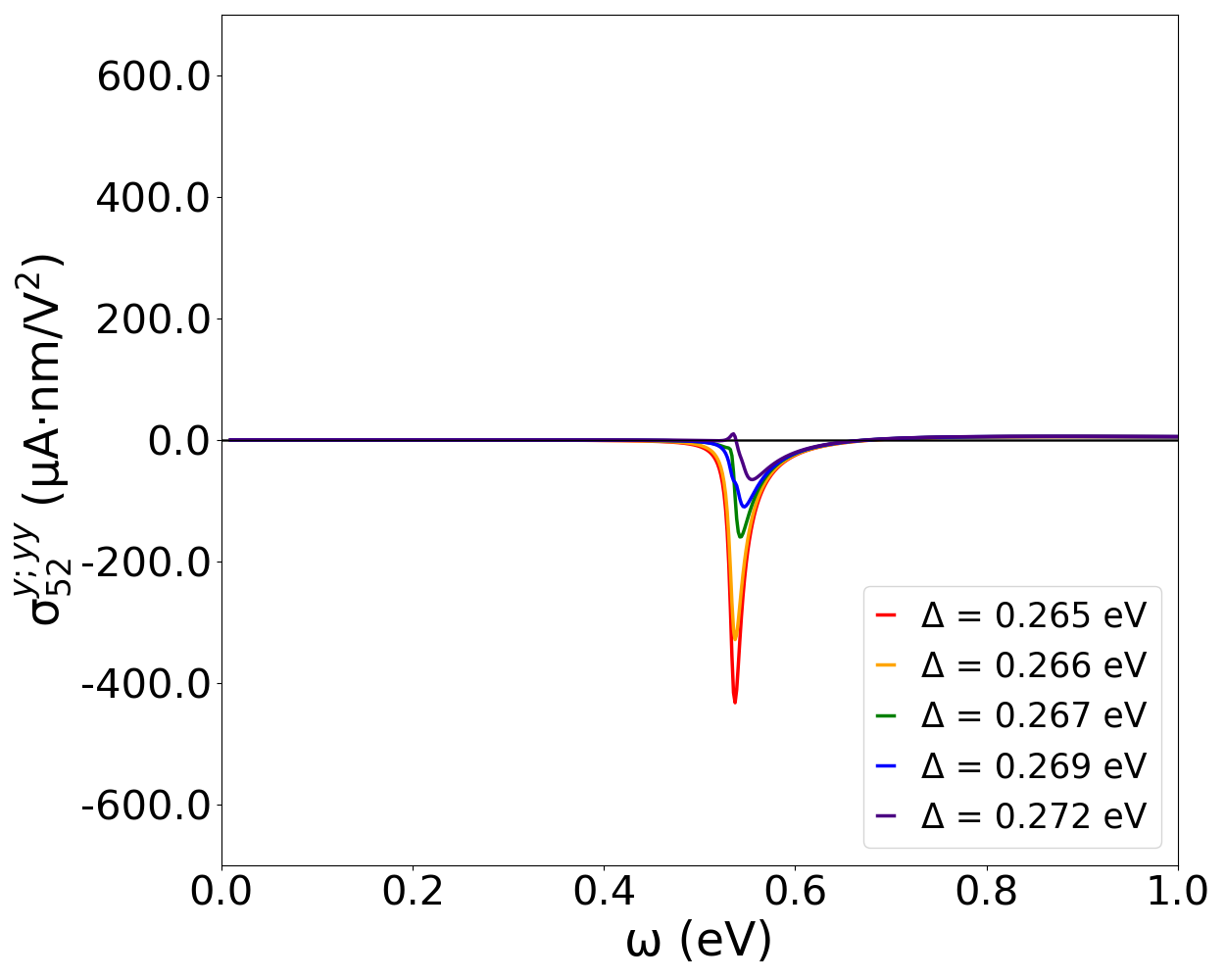}
    \caption{(top) The band-resolved conductivity $\sigma^{y;yy}_{43}$ contains the peak located at $\omega \approx 0.2\,\text{eV}$. The peak located at $\omega \approx 0.5\,\text{eV}$ is maximized at $\Delta = 0.269,\,0.272\,\text{eV}$. $\sigma^{y;yy}_{52}$ (bottom) contains a negative peak at a similar value of $\omega$ whose magnitude is maximized at $\Delta = 0.265, \,0.266\,\text{eV}$. In the total conductivity (see Fig. \ref{conductivity-colors}a), the features observed at $\omega \approx 0.5 \,\text{eV}$ are thus dominated by the behavior of $\sigma^{y;yy}_{52}$ at $\Delta = 0.265, \, 0.266\,\text{eV}$ and by $\sigma^{y;yy}_{43}$ at $\Delta = 0.269,\,0.272\,\text{eV}$. }
    \label{fig:syyy43-52}
\end{figure}

We observe in Fig.\ref{figsigmaAB_ABC} that the conductivity of AB bilayer graphene and ABC trilayer graphene share a similar peak located around $\omega \approx 0.2\,\text{eV}$ for $\Delta=0.2\,\text{eV}$, with the conductivity of ABC trilayer graphene displaying additional features located in the $0.4\,\text{eV}\lesssim\omega \lesssim 0.7\,\text{eV}$ range. The position of this peak stays relatively fixed at at $\omega \approx0.2\,\text{eV}$ as $\Delta$ varies for ABC trilayer graphene while it varies from $\omega \approx 0.2-0.35\,\text{eV}$ in AB bilayer graphene. However, the conductivity of ABC trilayer graphene displays markedly different behavior from the AB bilayer graphene in the $0.4\,\text{eV}\lesssim \omega \lesssim 0.6\,\text{eV}$ frequency range. Over this frequency range the ABC trilayer graphene displays a sign change as $\Delta$ changes from 0.2 to 0.3 eV, while the AB bilayer does not.

The conductivity and band structure were studied at values of $\Delta$ between $0.2\,\text{eV}$ and $0.3\,\text{eV}$ to determine if features in the band structure corresponded to the sign change in the conductivity. It was observed that the conductivity changes negligibly as $\Delta$ increases from $0.265\,\text{eV}$ to $0.266\,\text{eV}$, with the band structure similarly displaying little change between the two values of $\Delta$. However, the conductivity at $\omega=0.5\,\text{eV}$ changes significantly as $\Delta$ is tuned from $0.266\,\text{eV}$ to $0.267\,\text{eV}$ and $0.269\,\text{eV}$. 

It can be seen in Fig.~\ref{conductivity-colors}a that the peak in the conductivity located at $\omega \approx 0.2\,\text{eV}$ varies negligibly with $\Delta$, while the additional features observed between $\omega = 0.4\,\text{eV}$ and $0.6\,\text{eV}$ evolve significantly as $\Delta$ varies from $0.265\,\text{eV}$ to $0.272\,\text{eV}$. Closer examination of the band pairs directly above and below the Fermi level, i.e, for bands 2-3 and 4-5 (see Fig.\ref{figbandstructure} for band labeling convention) reveals a gap closing between $\Delta = 0.266\,\text{eV}$ and $0.267\,\text{eV}$, as well as another band touching at $\Delta = 0.269-0.270\,\text{eV}$ as shown in the dashed inset of Fig.~\ref{conductivity-colors}(a).

The conductivity changes minimally at the first band touching at $\Delta \approx 0.266\,\text{eV}$, but changes sign as $\Delta$ is tuned through values past the first gap closure, approaching the second gap closure at $\Delta \approx 0.269\,\text{eV}$.  Therefore, the detailed changes of the nonlinear shift conductivity provides important information about band evolution with displacement field $\Delta$.  As we will show below, there is also information on the quantum geometric properties of the electronic wavefunctions in the evolution of the conductivity.

For both of these band pairs, as seen in Fig.~\ref{conductivity-colors}(a), the band structure along a $\Gamma-\text{K}^\prime-\text{K}$ path exhibits a band inversion as the displacement field is varied from $\Delta = 0.266\,\text{eV}$ to $0.269\,\text{eV}$, a detail not visible when examining the band structure along the $\Gamma-\text{K}^\prime-\text{M}$ path. This band inversion is expected to have a significant impact on the Berry curvature and other quantum geometric properties of ABC trilayer graphene~\cite{cayssol2021topological,Xiao2010}. Typically, Berry curvature (and also the quantum metric) tend to concentrate around bound touching (or closest approach points of bands) in the Brillouin zone. (See Fig.\ref{fig:qm}.) No such band inversion is observed in the band structure of AB bilayer graphene with the variation of $\Delta$ once the gap around Fermi energy is opened.

To further investigate the features noted in the conductivity for values of displacement field in the range $\Delta=0.265\,\text{eV}-0.272\,\text{eV}$, the band-resolved conductivity $\sigma_{nm}^{y;yy}(\omega)$ is computed using
\begin{equation}
    \sigma^{a;bc}_{nm}(\omega)= \frac{2g_s\pi e^3}{\hbar^2} \int \frac{dk^2}{(2\pi)^2}f_{nm}I_{nm}^{a;bc}\delta(\omega_{nm}-\omega)
\end{equation}
where $n,m$ are the indices for bands involved in the transition.

\begin{figure*}
    \centering
    \includegraphics[scale=0.275]{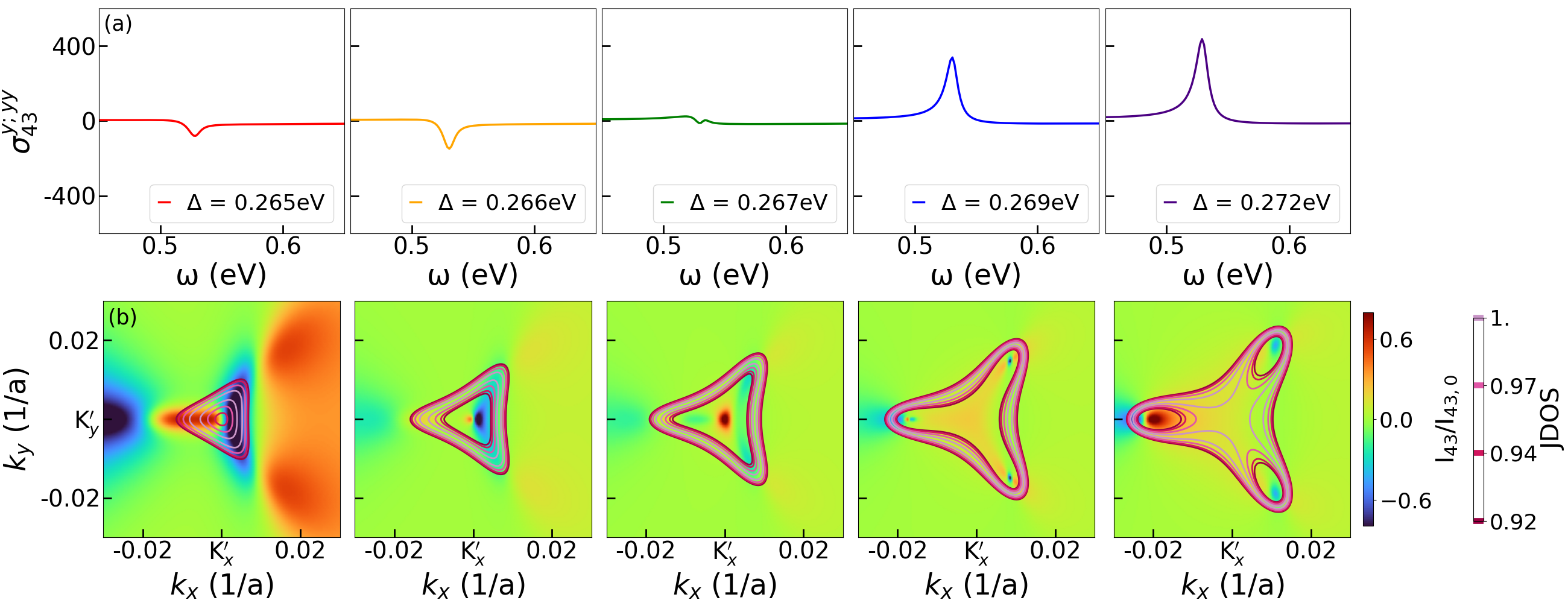}
    \includegraphics[scale=0.275]{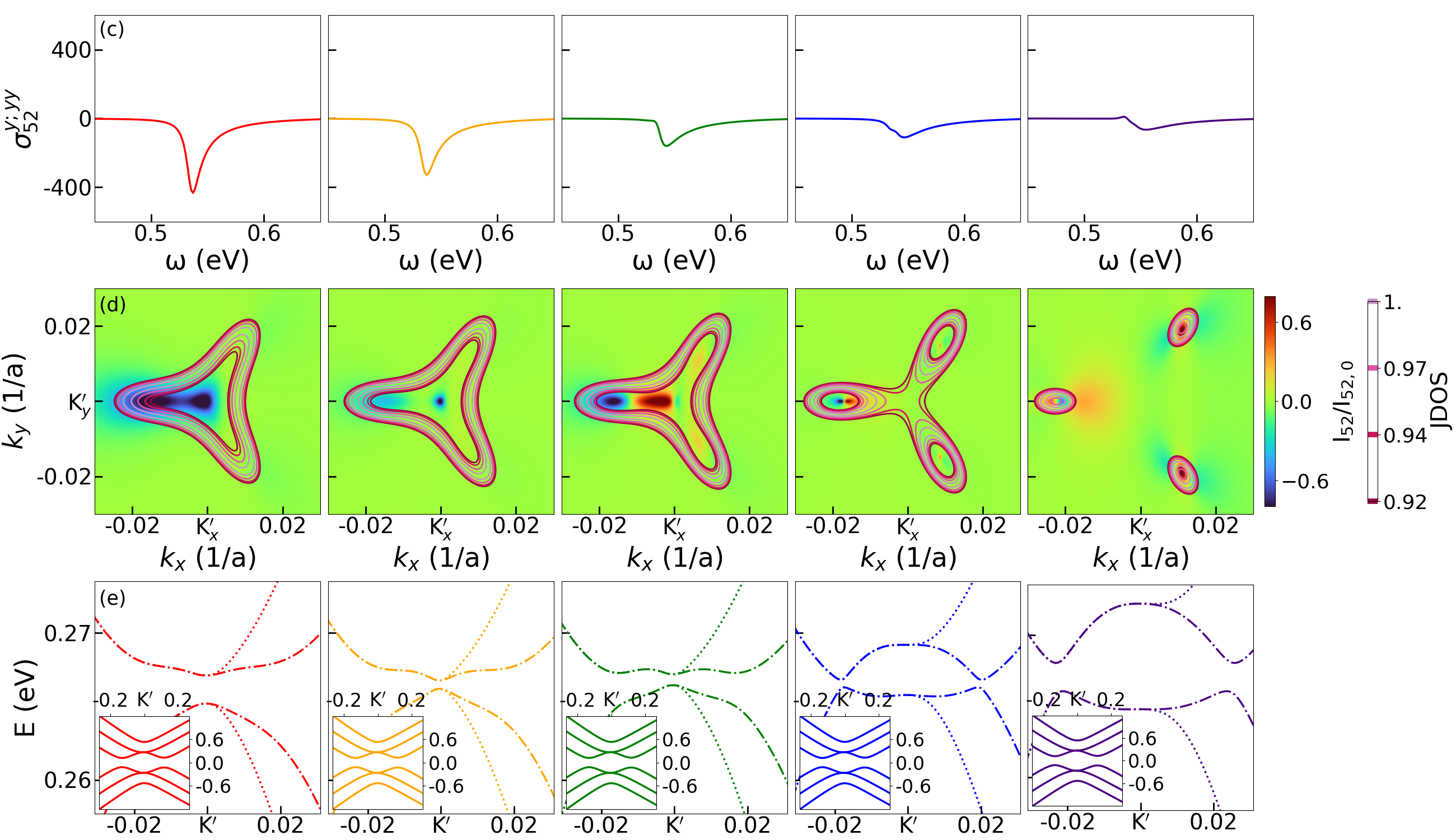}
    \caption{(a) The band-resolved conductivity $\sigma^{y;yy}_{43}(\omega)$ shows a sign change from negative to positive as $\Delta$ increases from $0.265\,\text{eV}$ to $0.272\,\text{eV}$. (b) The JDOS and band-resolved integrand $I_{43}(\mathbf{k})$, where $I_{43,0}$ is the maximum value of the integrand within the $\vb{k}$-space window surrounding $\text{K}^\prime$. (c) The band-resolved conductivity $\sigma^{y;yy}_{52}(\omega)$ shows a sign change from positive to negative as $\Delta$ increases from $0.265\,\text{eV}$ to $0.272\,\text{eV}$. (d) The JDOS and band-resolved integrand $I_{52}(\mathbf{k})$, where $I_{52,0}$ is the maximum value of the integrand within the $\vb{k}$-space window surrounding $\text{K}^\prime$. (e) Along the $\Gamma-\text{K}^\prime-\text{K}$ path through the FBZ (dash-dot), a band inversion occurs as $\Delta$ increases from $0.265\,\text{eV}$ to $0.272\,\text{eV}$.}
    \label{sigintband_1}
\end{figure*}

The band-resolved conductivity plotted in  Fig.~\ref{conductivity-colors}b reveals that the major features of the conductivity are contained in $\sigma^{y;yy}_{43}$, $\sigma^{y;yy}_{52}$,$\sigma^{y;yy}_{53}$ and $\sigma^{y;yy}_{42}$, with $\sigma^{y;yy}_{53}$ and $\sigma^{y;yy}_{42}$ identical due to the symmetry of the graphene band structure about the Fermi level. Contributions from other transitions $\sigma^{y;yy}_{41}=\sigma^{y;yy}_{63}$, $\sigma^{y;yy}_{51}=\sigma^{y;yy}_{62}$ and $\sigma^{y;yy}_{61}$ were determined to have negligible effect on the features observed in the total conductivity $\sigma^{y;yy}$.

The overall shape of the conductivity, including the feature common to both AB and ABC graphene observed at $\omega \approx 0.2\,\text{eV}$ for $\Delta \approx 0.2\,\text{eV}$ (see Fig.~\ref{figsigmaAB_ABC}, Fig.~\ref{fig:syyy43-52}), is contained in $\sigma^{y;yy}_{43}$ which indicates the contribution from transitions between valence and conduction bands. As the displacement field increases, the magnitude of the negative minimum observed in $\sigma^{y;yy}_{52}$ and $\sigma^{y;yy}_{42}=\sigma^{y;yy}_{53}$ can be seen to decrease, with this feature reaching a minimum amplitude in $\sigma^{y;yy}_{52}$ and becoming a positive maximum in $\sigma^{y;yy}_{43}$, $\sigma^{y;yy}_{53}$ and $\sigma^{y;yy}_{42}$ (see Fig.~\ref{conductivity-colors}b). This change in sign coincides with the gap closure between bands 2(4) and 3(5). Such gap closures are expected to alter the quantum geometric properties, and can thus significantly affect the shift-current conductivity studied here. The shift-current response is determined by the integrand $I_{nm}^{y;yy} (\mathbf{k})$ which is derived directly from the Hermitian connection~\cite{ahn2022riemannian} along with the JDOS which determine what regions of Brillouin zone can contribute at a given frequency. The zero-temperature JDOS, $\mathcal{J}_{nm}$ can be computed using 

\begin{equation}
    \mathcal{J}_{nm}(\omega) = \int \frac{dk^2}{(2\pi)^2}\delta(\omega_{nm}(\vb{k})-\omega). 
\end{equation}

Next we study the integrand $I_{nm}^{y;yy}(\mathbf{k})$  alongside the corresponding JDOS in a small window around $K^\prime$, near the band touchings, to determine the contribution of different $\vb{k}$-space regions to the total conductivity.  It can be seen in Fig.~\ref{sigintband_1}b and Fig.~\ref{sigintband_1}d  that the JDOS does not contain the $\vb{k}$-space points at which the bands touchings occur, but does contain the surrounding regions in $\vb{k}$-space and thus the associated features in the band structure of the system. This suggests that the sign change of the conductivity results from the changes in the shape of the band structure that occur at these values of $\Delta$, rather than the band touchings themselves, corresponding to the inversion of the band conductivity that occurs as $\Delta$ increases from $0.265\,\text{eV}$ to $0.272\,\text{eV}$. 

Both $\sigma_{43}^{y;yy}$ and $\sigma_{52}^{y;yy}$ display a sign change from positive to negative, but the magnitude of $\sigma^{y;yy}_{43}$ is positive at its greatest magnitude while the reverse is true for $\sigma^{y;yy}_{52}$. Examining $\sigma^{y;yy}_{43}$ in more detail, we can see the evolution of the integrand with varying $\Delta$, as well as the features picked up by the JDOS  $\mathcal{J}_{43}(\mathbf{k},\omega)$ evaluated at $\omega = 0.529\,\text{eV}$ (selected by evaluating the value of $\omega$ at which $\sigma^{y;yy}_{43}$ achieves its maximum amplitude at $\Delta = 0.269\,\text{eV}$). 

The region of the band-resolved integrand $I^{y;yy}_{43}$ captured by the JDOS $\mathcal{J}_{43}(\mathbf{k},\omega)$ in Fig.~\ref{sigintband_1}b has both negative and positive regions for $\Delta = 0.265\,\text{eV}$, with the negative region dominating the total band-resolved conductivity, leading to the overall negative amplitude. The magnitude of the positive region decreases as $\Delta$ increases to $0.266\,\text{eV}$. At $\Delta = 0.267\,\text{eV}$, a positive region with a large magnitude develops at $K^\prime$ but is not captured by the JDOS. However, the magnitude of the negative region captured by the JDOS decreases. The magnitude of the negative peak seen in the band-resolved conductivity likewise decreases (see Fig.~\ref{sigintband_1}a). 

At $\Delta = 0.269\,\text{eV}$, the total band-resolved conductivity changes both in sign and magnitude, displaying a positive peak located at $\omega \approx 0.53 \,\text{eV}$. The integrand can be seen to develop three negative regions at the band-touching locations in the first Brillouin zone. The JDOS does not capture these regions of the integrand, but rather the surrounding positive region. This peak then grows in magnitude as $\Delta$ increases to $0.272 \,\text{eV}$, at which point a defined positive region appears in the integrand, the outermost portion of which is captured by the JDOS. This sign change correlates with the band inversion observed at $\Delta = 0.269\,\text{eV}$ (see Fig.~\ref{sigintband_1}e), with the magnitude of the curvature of band 4 increasing as $\Delta$ increases from 0.269 to 0.272 eV, corresponding to the increasing magnitude of the total band-resolved conductivity. Given the symmetry of the system band structure, when band 4 reverses from negative curvature to positive curvature, band 3 reverses from positive to negative curvature at the point of the sign change in the total band-resolved conductivity. 

The peak observed in the total band-resolved conductivity $\sigma^{y;yy}_{52}$ shows changes contrasting with those observed in $\sigma^{y;yy}_{43}$, displaying maximum amplitude at the negative peak occurring at $\Delta = 0.265\,\text{eV}$ and $\omega = 0.537\,\text{eV}$, which determined the value of $\omega$ at which the JDOS $\mathcal{J}_{52}(\mathbf{k},\omega)$ was evaluated. The integrand shows a clear negative region at $K^\prime$, but this region is not selected by the JDOS. Another region that is not as sharply defined is instead picked up by the JDOS (see Fig.~\ref{sigintband_1}d). At $\Delta = 0.267\,\text{eV}$, the magnitude of the negative peak markedly decreases (see Fig.~\ref{sigintband_1}c) as a positive region develops in the integrand, the edges of which are captured by the JDOS (Fig.~\ref{sigintband_1}d). The spans of both the positive and negative regions of the integrand are greatly compressed at $\Delta = 0.267\,\text{eV}$ with the JDOS enclosing the regions without capturing the maximal positive or negative portions of the integrand. The total band-resolved conductivity in Fig.~\ref{sigintband_1}d decreases in magnitude correspondingly. The JDOS also does not follow the $\Gamma-K^\prime-K$ line of symmetry along which the band inversion is observed in Fig.~\ref{sigintband_1}e, and no sign change is observed in the $\sigma^{y;yy}_{52}$ component of the band-resolved conductivity (Fig.~\ref{sigintband_1}c). 

These features were further studied through analysis of the eigenvector projections via calculation of the probability amplitude for each of the lattice sites (see Appendix \ref{eigapp}), as well as through the lens of geometric quantities $Q_{ba}^{n}$ and $C^{nm}_{cba}$ (see Appendix \ref{Qabnapp} and \ref{Cyyyapp}). The probability amplitudes were calculated along both the $\Gamma-K^\prime-K$ and $\Gamma-K^\prime-M$ symmetry lines in the FBZ. Tuning of the parameter $\Delta$ had a marked impact on the eigenvector projections. The probability amplitude corresponding to each site was clearly maximized for band pairs either directly above or below the Fermi energy with the magnitude and curvature changing dramatically with the variation of $\Delta$ from 0.265 to 0.267 eV (see Fig.~\ref{EVpA1B1}, Fig.~\ref{EVpA2B2}, and Fig.~\ref{EVpA3B3}). We were ultimately unable to precisely link the features in the eigenvectors to the changes observed in the total conductivity, band-resolved conductivity, or band-resolved integral. The calculation of the Berry curvature and quantum metric (see Appendix \ref{Qabnapp}) likewise revealed that both geometric quantities are significantly affected by the variation of $\Delta$, with the $\vb{k}$-space extrema shifting from a single extremum located at $K^\prime$ to three separate extrema located along $\Gamma-K^\prime-M$ symmetry lines in the FBZ (see Fig.~\ref{fig:qm}). The single maxima observed at $K^\prime$ in the quantum metric develops into three distinct maxima along this symmetry line (see Fig.~\ref{fig:qm}a) as $\Delta$ changes from 0.267 to 0.269 eV. In the Berry curvature, a positive (negative) extrema changes sign at $\Delta = 0.267\,\text{eV}$ and then becomes three distinct negative (positive) extrema along the high-symmetry path in the FBZ at $\Delta = 0.269\,\text{eV}$ (see Fig.~\ref{fig:qm}b).

The metric connection and symplectic connection, $\Gamma^{nm}_{bca}$ and $\tilde\Gamma^{nm}_{bca}$ were calculated from the real and imaginary parts of the Hermitian connection, respectively (see Appendix \ref{Cyyyapp}). The metric connection displays complex features that also evolve markedly with $\Delta$. Much like the quantum metric and Berry curvature, the metric connection and symplectic connection develop three distinct extrema along the $\Gamma-K^\prime-K$ symmetry line in the FBZ at $\Delta = 0.267\,\text{eV}$, along with various sign changes (see Fig.~\ref{fig:hc}). Although features corresponding to the tuning of $\Delta$ were clearly observed in these quantities, a concise correlation between the changes observed in the conductivity and these geometric quantities could not be determined.

\section{Conclusions}
\label{sec:conc}

Rhombohedral trilayer graphene has recently emerged as a strong alternative to twisted bilayer graphene systems for studying the quantum geometry and correlation driven physics~\cite{yacoby2011tri}. Photocurrent response serve as a highly reliable and useful knob to probe  these aspects in quantum materials~\cite{ma2023photocurrent,kumar2024terahertzphotocurrentprobequantum}. In particular, bilayer graphene and twisted multilayer graphene have been predicted to exhibit highly-tunable large bulk photovoltaic effect arising from the shift-current conductivity~\cite{xiong2021atomic,Kaplan2022TBG,chaudhary2022shift,chen2024enhancing}.

In this work, we showed that trilayer graphene also exhibit a large second-order DC response originating from the quantum geometric features of the system. We calculated the low-frequency ($\omega <1\,\text{eV}$) shift current response of ABC trilayer graphene with inversion symmetry broken via a displacement field described by the parameter $\Delta$. A comparison between the shift-current conductivity $\sigma^{y;yy}(\omega)$ of AB bilayer graphene and ABC trilayer graphene revealed that in the low-frequency regime, the shape of the conductivity at $\omega \approx 0.2-0.4\,\text{eV}$ are similar between the two systems. However, at $\omega\approx 0.5\,\text{eV}$, the conductivity of ABC trilayer graphene displays additional features that change significantly with the tuning of the displacement field. 

These features mainly arise due to additional bands in the trilayer case that undergo band inversion with displacement field unlike the bilayer case where the uppermost and lowermost bands do not experience a gap closing with the displacement field.  A particular feature in the shift current conductivity was observed to exhibit a sign change at this band inversion point. This was also accompanied by a significant change in the behaviour of the Hermitian connection and shift-current conductivity which indicates the quantum geometric nature of these nonlinear optical responses. A significant enhancement in conductivity around these points can also be understood in terms of the contributions from virtual transition terms in the shift-current conductivity which increases with decreased energy separation between bands involved in virtual transitions~\cite{Parker19,chen2024enhancing}.

In addition to large quantum geometric features, ABC trilayer graphene systems are also known to exhibit strong signatures of electron-electron interactions \cite{yacoby2011tri,zhou2021superconductivity,Huang2023,Koh2024,PhysRevB.109.L060409,Ghazaryan2023}. Such interaction effects can often lead to correlated states with spontaneously broken symmetries~\cite{jung2013,zhou2021half,Huang2023}.  Nonlinear optical responses are highly sensitive to broken spatial and time-reversal symmetries. The consequences of these interactions on shift-current response is an important topic for future study where a variation with chemical potential may reveal important signatures of the interplay between interactions and quantum geometry. Moreover, even in untwisted layered graphene systems, hBN substrate can give rise to superlattice potentials, which modify the band structure and quantum geometric properties significantly~\cite{Patri2023,Ghorashi2023PRB,Ghorashi2023PRL,PhysRevB.109.195406}. These moire effects should give rise to characteristic signatures in nonlinear optical responses.
\section{Acknowledgements}
We gratefully acknowledge funding from the National Science Foundation through the
Center for Dynamics and Control of Materials: an
NSF MRSEC under Cooperative Agreement No. DMR-
1720595 and DMR-2114825. G.A.F. acknowledges additional support from the Alexander von Humboldt Foundation. SC would also like to acknowledge discussions with Sihan Chen, Cyprian Lewandowski, and Roshan K. Kumar.

\appendix

\section{Eigenvector projections}\label{eigapp}

The eigenvector projections from the six sites in unit cell of the ABC trilayer graphene system were calculated to see if any direct contributions to the conductivity could be identified. We begin with 
\begin{equation}
    H(\vb{k})|n\rangle = E_n |n\rangle,
\end{equation}
with the eigenfunctions $|n\rangle$ calculated using the basis $\{\ket{A_1},\ket{B_1},\ket{A_2},\ket{B_2},\ket{A_3},\ket{B_3}\}$ of the ABC trilayer graphene.

\begin{figure}[ht!]
    \centering
    \includegraphics[scale=0.275]{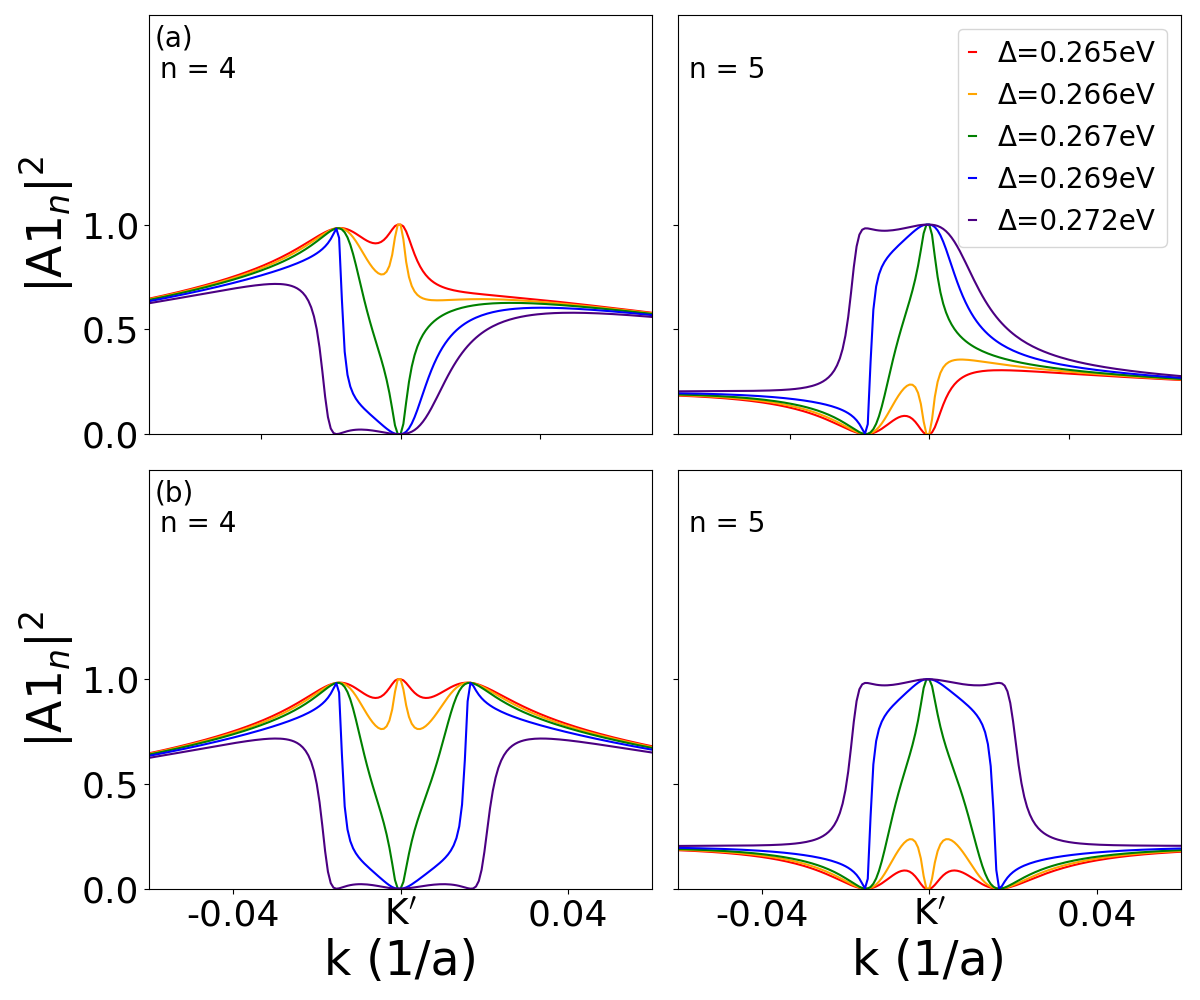}
    \includegraphics[scale=0.275]{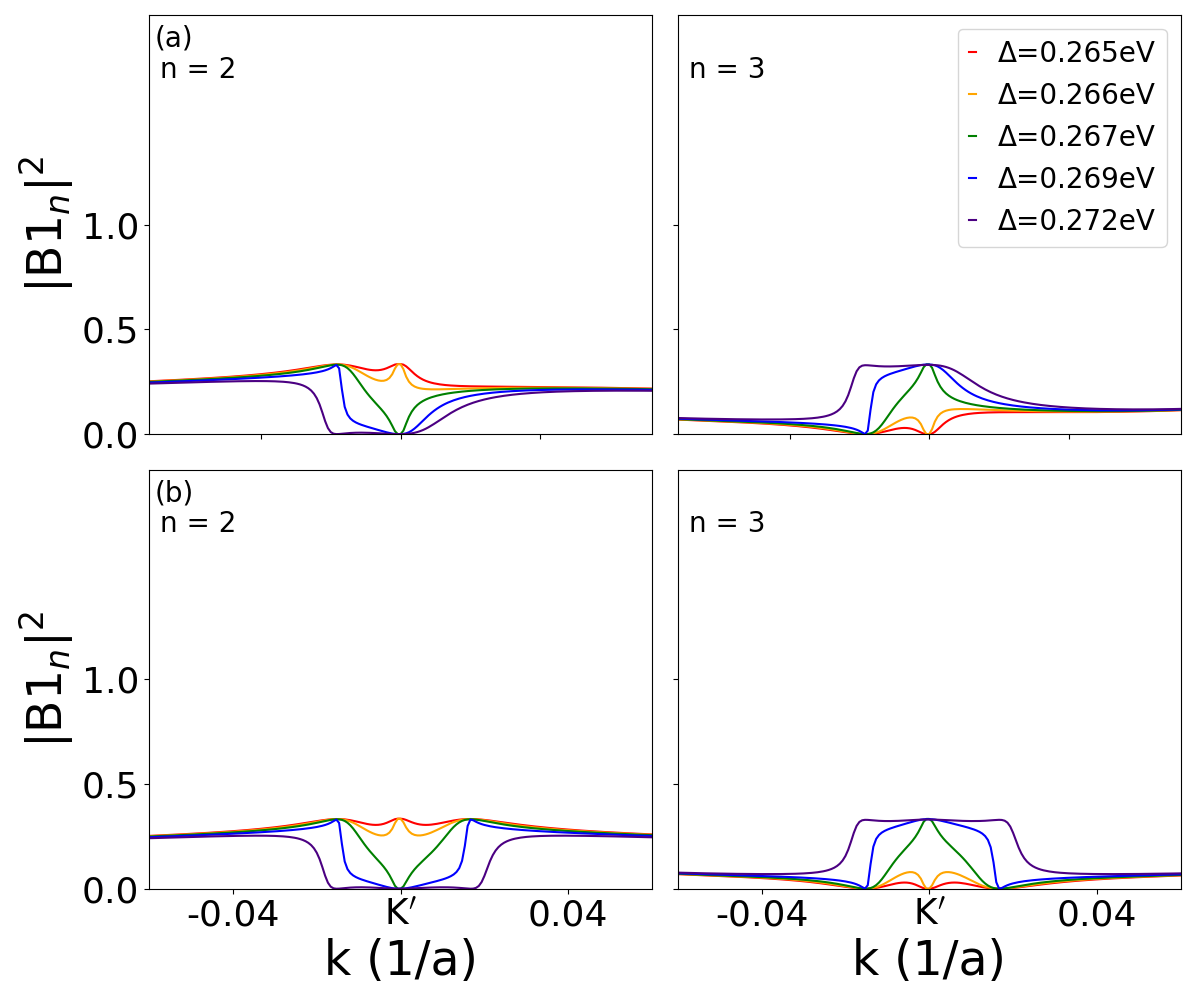}
    \caption{Eigenvector projections corresponding to the $A_1$ and $B_1$ sites in ABC trilayer graphene along the (a) $\Gamma-K^\prime-M$ and (b) $\Gamma-K^\prime-K$ symmetry lines through the FBZ. It can be seen that the $A_1$ eigenvector projection is maximized at $\Delta = 0.265 \,\text{eV}$ and decreases as $\Delta$ increases for $n$ = 4, while the reverse is true for $n = 5$. The same trend can be observed in the eigenvector projections corresponding to the site $B_1$. Furthermore, the $A_1$ eigenvector projections have the largest magnitude of the $A$-site projections, while the reverse is again true for the $B_1$ eigenvector projections, which represents the smallest magnitude of the $B$-site projections.}
    \label{EVpA1B1}
\end{figure}

\begin{figure}[ht!]
    \centering
    \includegraphics[scale=0.27]{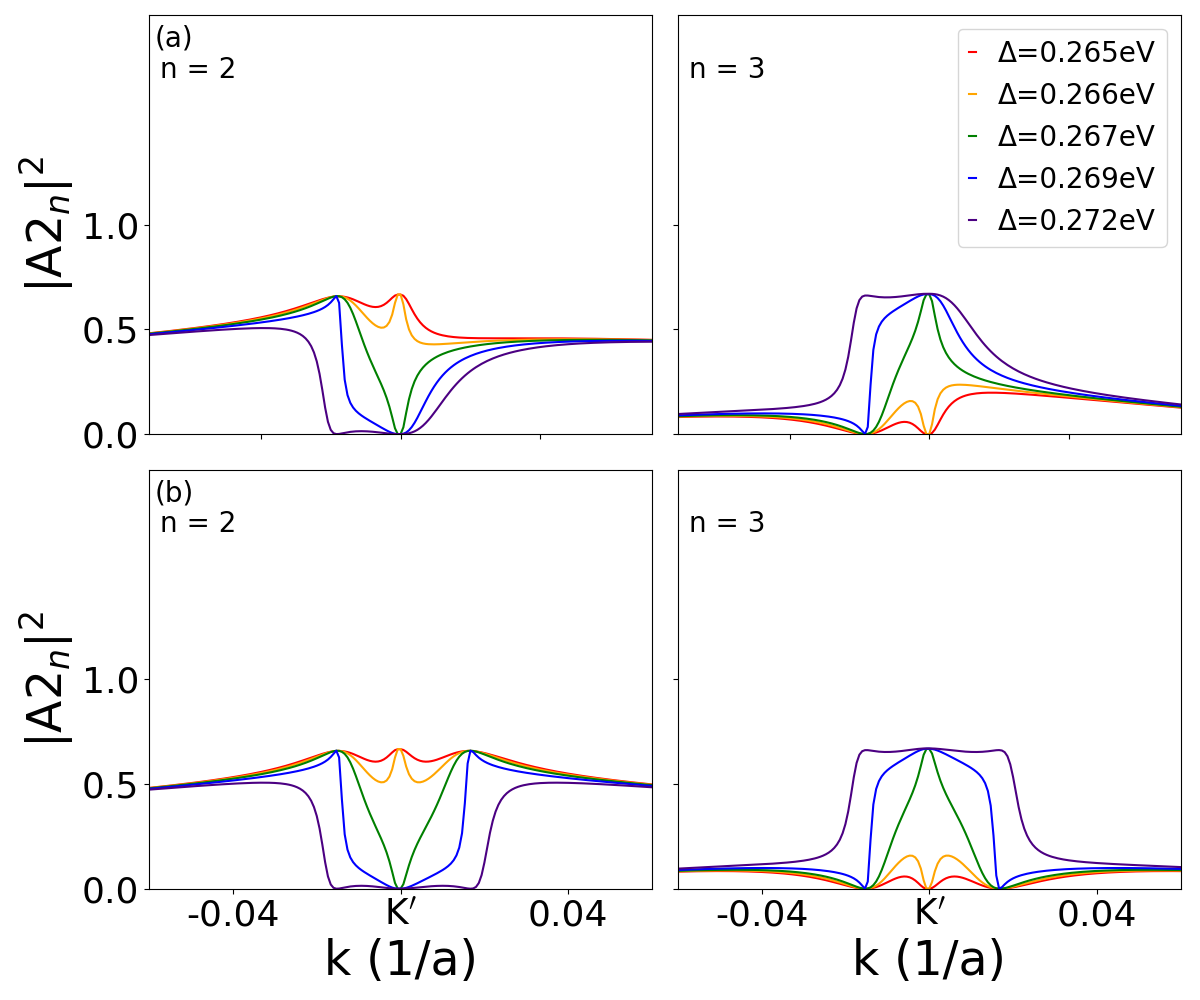}
    \includegraphics[scale=0.27]{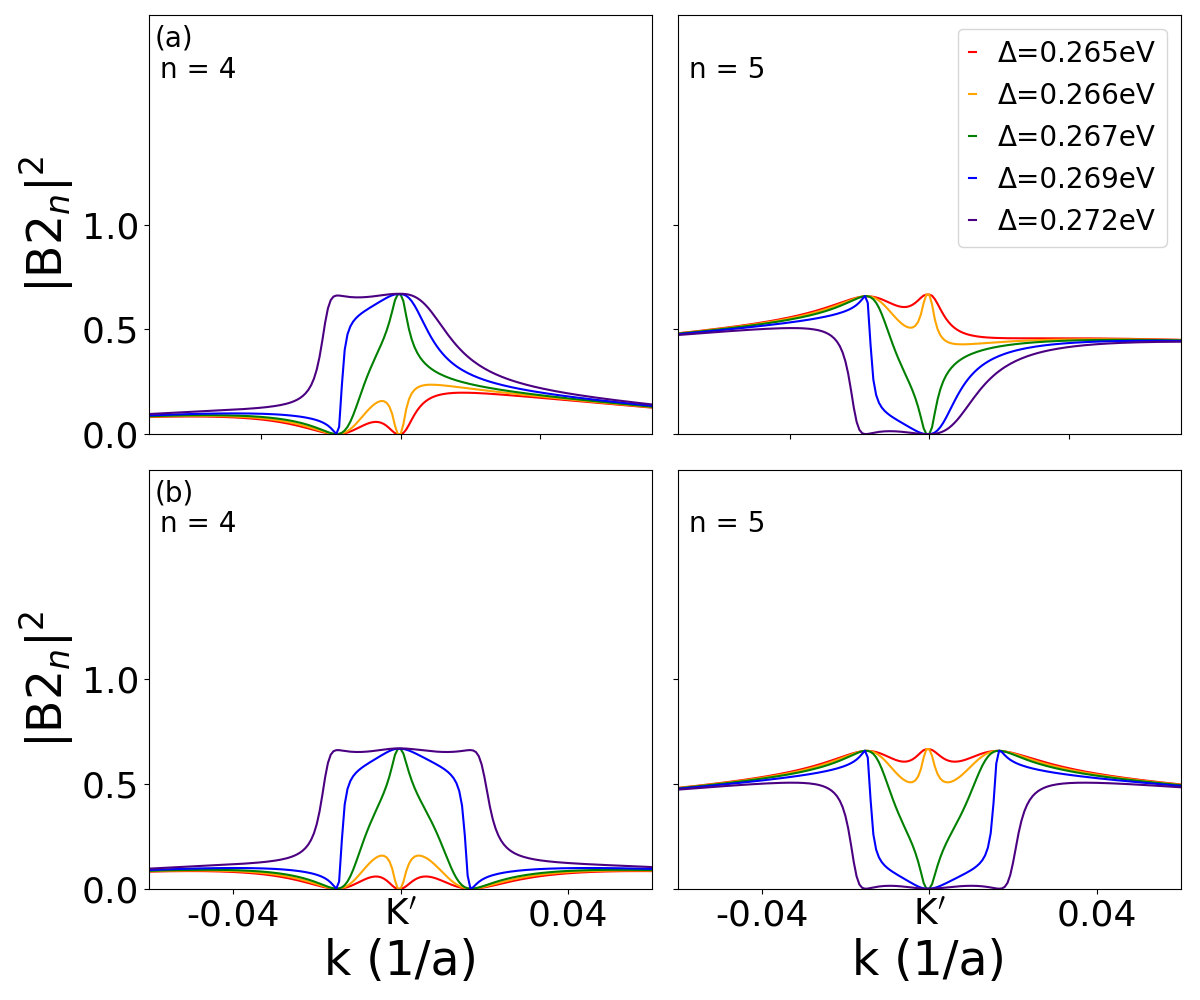}
    \caption{Eigenstate projections corresponding to $A_2$ and $B_2$ sites of ABC trilayer graphene along the (a) $\Gamma-K^\prime-M$ and (b) $\Gamma-K^\prime-K$ symmetry lines through the FBZ. $A_2$ is maximized for bands 2 and 3, while $B_2$ is maximized for bands 4 and 5. Unlike the $A_1$ projections, the $A_2$ projections are maximal for $n =$ 2, 3 rather than 4, 5. The same contrast is true for the $B_2$ eigenvector projections.}
    \label{EVpA2B2}
\end{figure}

\begin{figure}[ht!]
    \centering
    \includegraphics[scale=0.275]{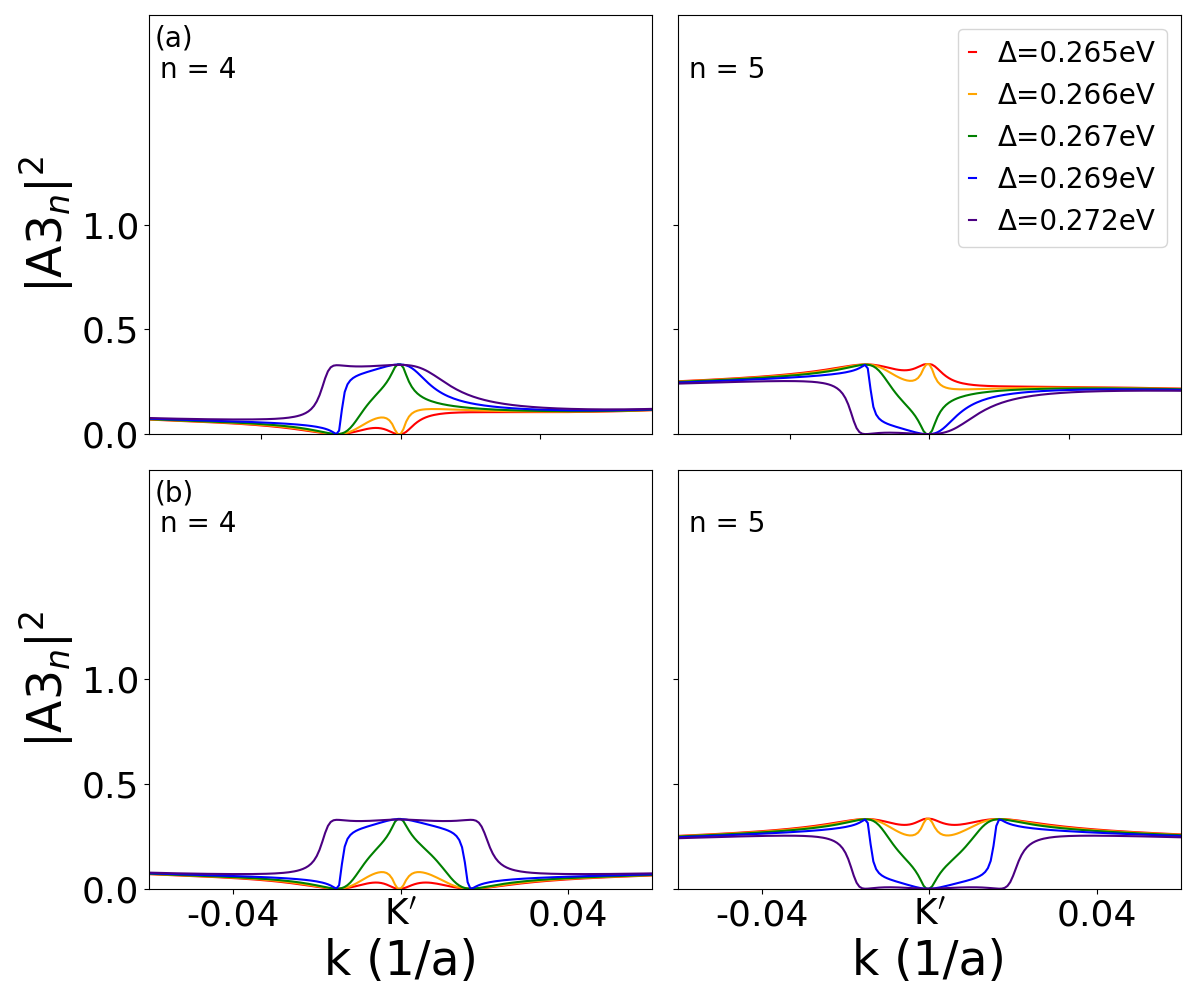}
    \includegraphics[scale=0.275]{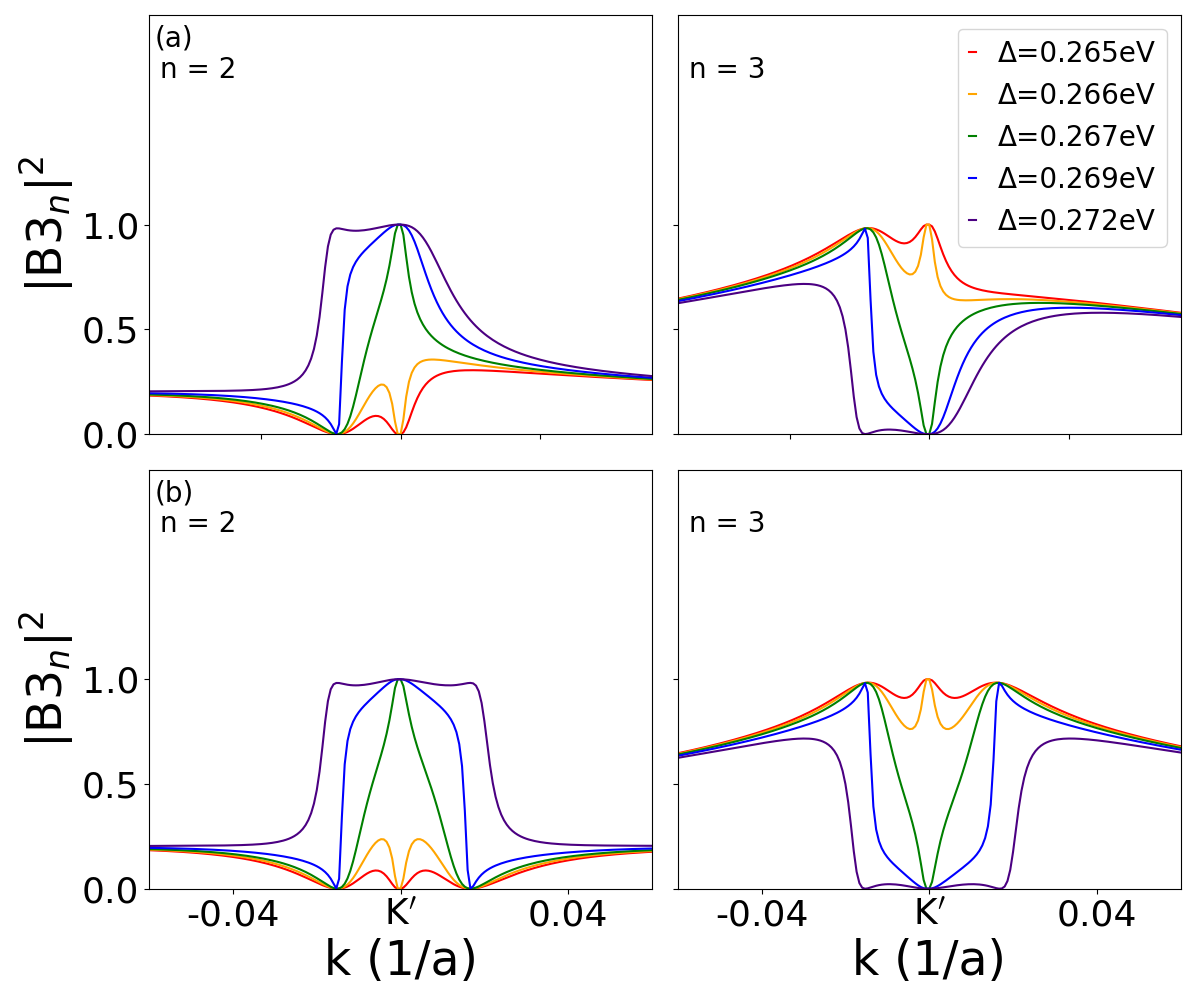}
    \caption{Eigenvector projections corresponding to $A_3$ and $B_3$ sites in ABC trilayer graphene along the (a) $\Gamma-K^\prime-M$ and (b) $\Gamma-K^\prime-K$ symmetry lines through the FBZ. The $A_3$ projections follow a pattern opposite to $A_1$ but for the same values of $n$, with the projections maximal at $\Delta =$0.272 eV for $n = 4$ and at $\Delta = 0.265\,\text{eV}$ for $n = 5$.}
    \label{EVpA3B3}
\end{figure}

The  projections of the $\ket{n}$ eigenvector on a given sublattice of the $i^{th}$ layer is denoted by 
\begin{equation}
    |A_{i,n}|^2 = |\bra{n}\ket{A_i}|^2.
\end{equation}
Eigenvector projections from the six sites in the Hamiltonian of ABC trilayer graphene did not yield any clear insight into the origin of the sign change observed in the integrand (as described in the main text), although significant changes in the magnitude of each projection as $\Delta$ was varied from $0.265-0.272\,\text{eV}$.  This can be seen in Figs.~\ref{EVpA1B1}-\ref{EVpA3B3}.

One can observe that the magnitude of $A$-site projections decrease from $A_1$ to $A_3$, while $B$-site projections increase from $B_1$ to $B_3$. It can also be seen that if the eigenvector projections corresponding to an $A$-site of a given layer of the trilayer graphene lattice have the largest contributions from the band pair directly above (below) the Fermi level ($E = 0 \,\text{eV}$), then the eigenvector projections corresponding to the $B$-site of that layer will have the largest contributions from the band pairs directly below (above) the Fermi level. 
Further correlations can be seen between the variation of $\Delta$ and the magnitudes and geometric features of the eigenvector projections, most noticeably in the shape of the eigenvector projections as $\Delta$ increases from $0.265\,\text{eV}$ to $0.272\,\text{eV}$. For all sites, $\Delta = 0.265 \,\text{eV}$ corresponds to a maximum (minimum) with a local maximum (minimum) located at $K^\prime$. At $\Delta = 0.272\,\text{eV}$, this extremum becomes a minimum (maximum) with local minima (maxima) located offset from $K^\prime$ along the $\Gamma-K^\prime-K$ symmetry line through the FBZ (see Figs.~\ref{EVpA1B1}, \ref{EVpA2B2}, \ref{EVpA3B3}). However, studying these numerous changes in the eigenvector projections did not yield any clear insight as to the origin of the sign change in the conductivity. 

\section{Berry curvature and Quantum Metric}\label{Qabnapp}

The Berry curvature and quantum metric were computed from the quantum geometric tensor, given by 
\begin{equation}
    Q_{ab}^ n = \sum\limits_{m\neq n} r^a_{nm}r^b_{mn} \equiv g_{ab}^n - iF_{ab}^n/2
\end{equation}
with  its real and imaginary parts corresponding to the quantum metric $g_{ab}$ and the Berry curvature $F_{ab}$, respectively. As in the eigenvector projections discussed in Appendix A, the Berry curvature and quantum metric both displayed marked changes with tuning of the parameter $\Delta$, but did not provide a clear explanation for the origin of the sign change in the conductivity. 
\begin{figure*}[ht!]
    \centering
    \subfloat[]{\includegraphics[scale=0.27]{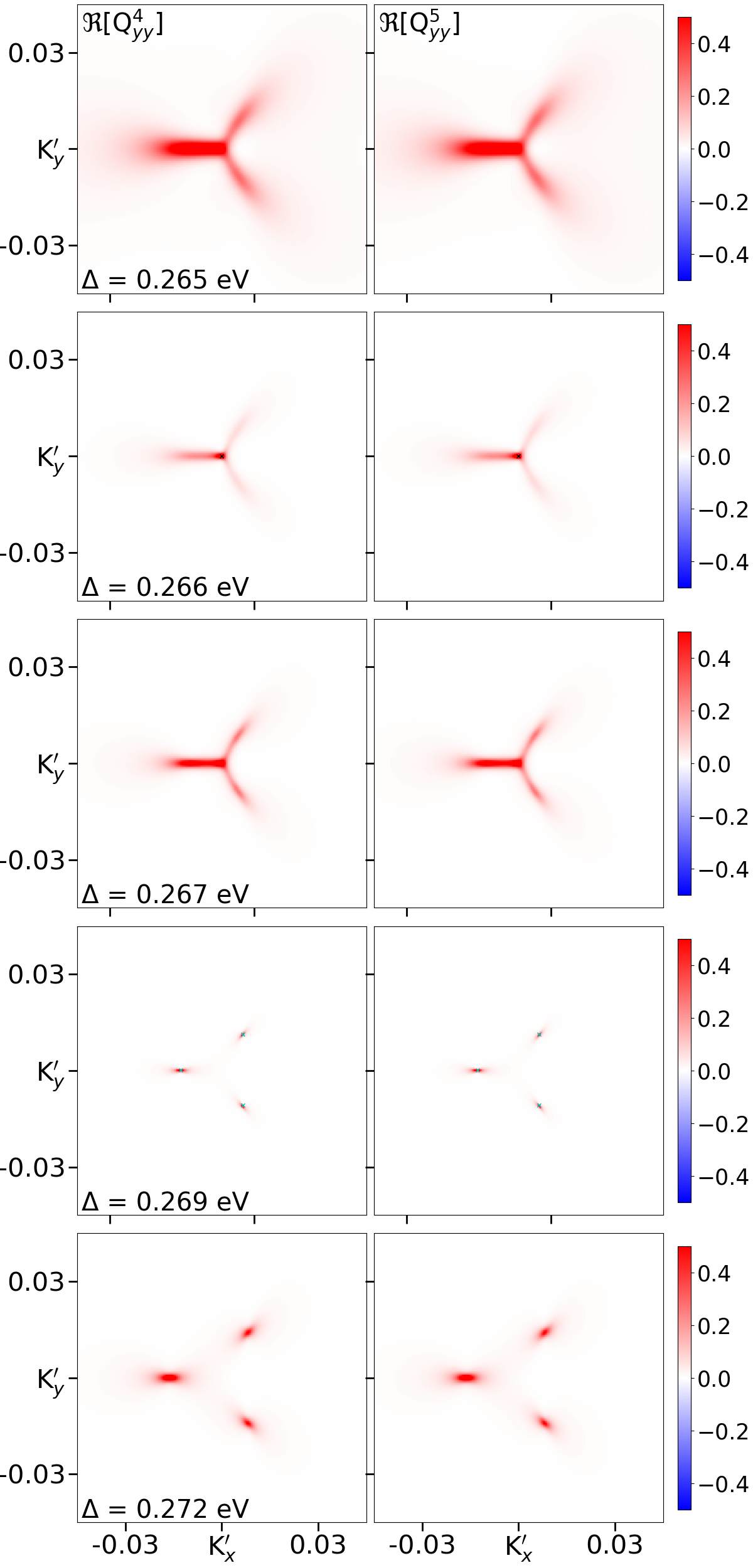}}
    \hspace{0.5cm}
    \subfloat[]{\includegraphics[scale=0.27]{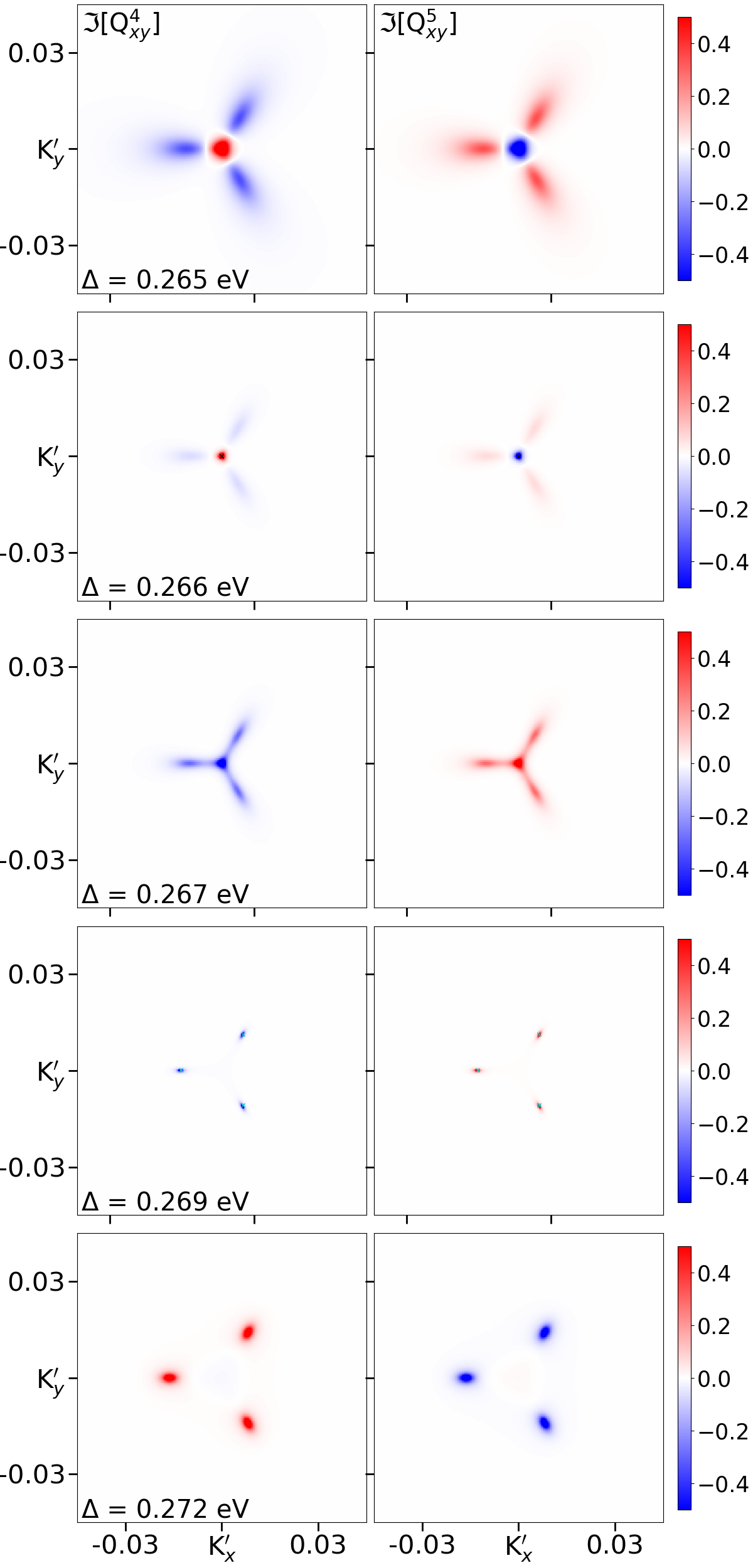}}
    \caption{The (a) quantum metric and (b) Berry curvature at $\Delta = 0.265,0.266,0.267,0.269$ and $0.272\,\text{eV}$. (a) The quantum metric develops three distinct maxima along the $\Gamma-K^\prime-M$ symmetry line through the FBZ at $\Delta = 0.269\,\text{eV}$. For the small $\vb{k}$-space window studied in the FBZ, the quantum metrics corresponding to bands 4 and 5 are identical, with additional features visible in the $\Re[Q^4_{yy}]$ at larger length scales. (b) The Berry curvature has a positive (negative) extremum located at $K^\prime$ at $\Delta = 0.265\,\text{eV}$ which changes sign at $\Delta = 0.267\,\text{eV}$. At  $\Delta = 0.269\,\text{eV}$, the Berry curvature displays three distinct positive (negative) extrema located along the $\Gamma-K^\prime-M$ symmetry line in the FBZ.}
    \label{fig:qm}
\end{figure*}
It can be seen in Fig.~\ref{fig:qm}a that the quantum metric corresponding to bands 4 and 5 changes drastically with the variation of $\Delta$, with a single maximum located at $K^\prime$ for $\Delta = 0.266\,\text{eV}$ morphing into three separate maxima along $\Gamma-K^\prime-K$ symmetry lines in the FBZ.

The Berry curvature undergoes a clear evolution as $\Delta$ is varied as well. At $\Delta = 0.265\,\text{eV}$, the Berry curvature displays a clear positive (negative) maximum surrounded by three negative (positive) regions for the Berry curvature corresponding to band 4 (5). As $\Delta$ increases, the $\vb{k}$-space area of the positive (negative) region decreases and the surrounding negative (positive) regions decrease in amplitude. At $\Delta = 0.267\,\text{eV}$, the positive (negative) region changes sign and merges with the surrounding regions of opposite sign. Finally, at $\Delta = 0.269\,\text{eV}$, the regions separate into three clear negative (positive) extrema surrounding a faint negative (positive) region.

\section{Hermitian Connection and $\sigma^{a;bc}$}\label{Cyyyapp}
\begin{figure*}[ht!]
    \centering
    \subfloat[]{\includegraphics[scale=0.27]{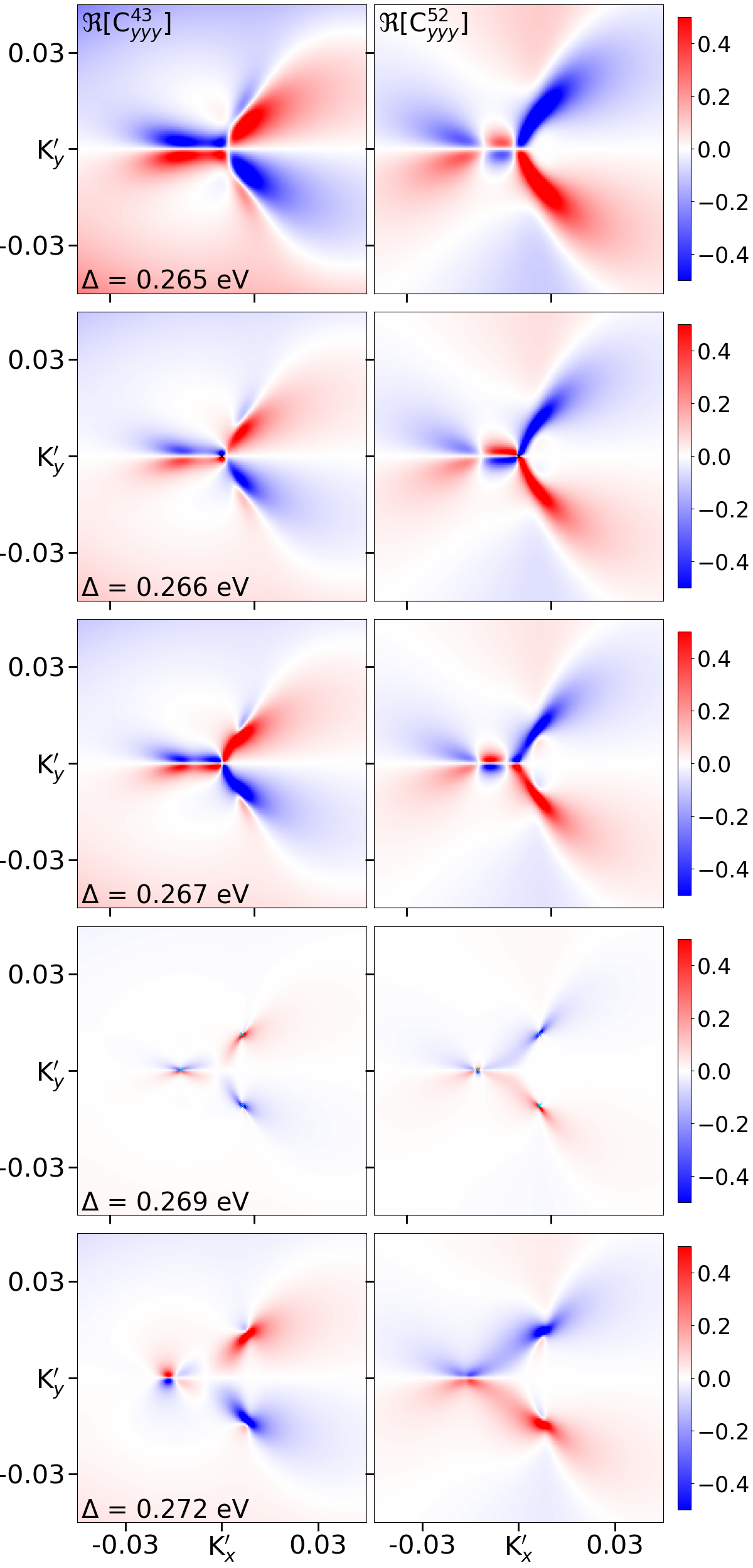}}
    \hspace{0.5cm}
    \subfloat[]{\includegraphics[scale=0.27]{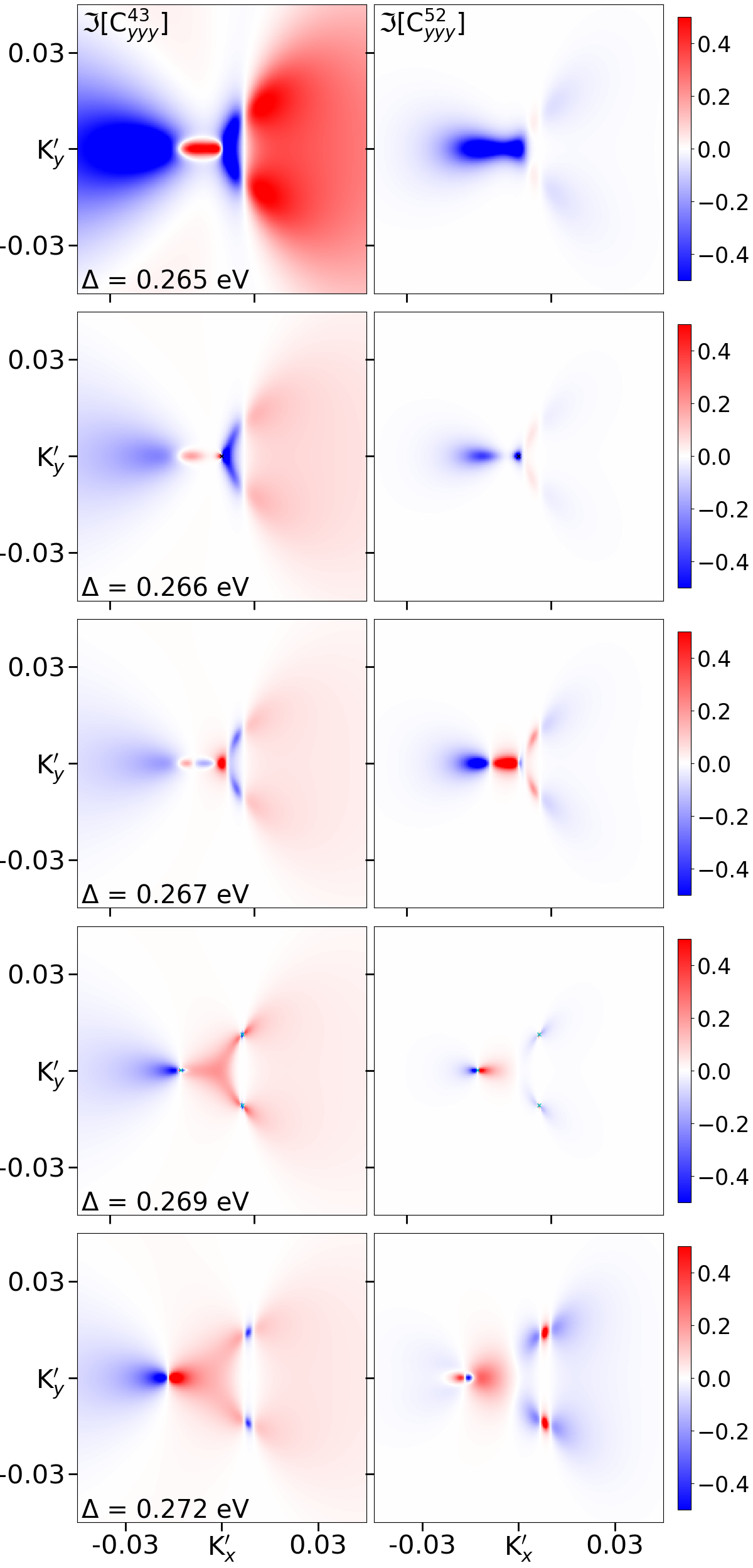}}
    \caption{The (a) real and (b) imaginary parts of the Hermitian connection for $(n,m)= (4,3)$ and $(5,2)$ (a) The metric connection corresponding to (n,m) = (4,3) shows alternative negative and positive regions with a nodal line along $k_y = 0$. These regions compress into nodes (antinodes) at $\Delta = 0.269\,\text{eV}$ and a sign change can be observed at $\Delta = 0.272\,\text{eV}$. (b) As $\Delta$ changes from 0.265 to 0.272 eV, two distinct negative (positive) extrema develop offset from $K^\prime$ at $k_y \neq 0$ along the $\Gamma-K^\prime-K$ symmetry line in the FBZ. At $k_y$ = 0, both a clear node and an antinode emerge.}
    \label{fig:hc}
\end{figure*}
The Hermitian connection is given by 
\begin{equation}
    C_{abc} = \sum\limits_{n \neq m} r^b_{nm}r^a_{mn;c} \equiv \Gamma_{bca} - i\tilde{\Gamma}_{bca}
\end{equation} 
where $\Gamma_{bca}$, $\tilde{\Gamma}_{bca}$ are the metric connection and symplectic connection, respectively. Defining the Hermitian connection as given is intended to assign a precise geometric meaning to the generalized derivative $r^b_{nm;a}$ \cite{ahn2022riemannian}.

The Hermitian connection can be used to rewrite the equation for the shift conductivity as
\begin{equation}
\sigma^{a;bc} = -\frac{\pi e^3}{2\hbar^2} \int\frac{dk^2}{(2\pi)^2} \sum\limits_{m,n} \delta(\omega - \omega_{mn})f_{nm}i(C^{mn}_{cab}-(C^{mn}_{bac})^*).
\end{equation}

The real and imaginary parts of the Hermitian connection both show features alongside the changes observed in the band structure. Along the $\Gamma-K^\prime-K$ symmetry line, the same line along which the band inversion is observed, the location of extrema and the signs of various regions of the metric connection change with $\Delta$, with the clearest development being the formation of distinct nodal and antinodal regions along the $\Gamma-K^\prime-K$ symmetry line (Fig.~\ref{fig:hc}a). The symplectic connection also markedly evolves with the variation of $\Delta$, with clear nodal and antinodal points developing along the $\Gamma-K^\prime-K$ symmetry line (Fig.~\ref{fig:hc}b).

Although the metric connection and Hermitian connection both significantly changed with the tuning of $\Delta$, the distinct features could not be concisely expressed in terms of simple sign changes or other easily quantifiable measures that could be correlated with the features noted in the conductivity. The imaginary part of the Hermitian connection is directly connected to the integrand used to compute the shift-current conductivity, but does not independently reveal characteristics of the conductivity without additional information from the other expressions involved in the integrand. 

\newpage
\bibliography{ref}

\end{document}